\documentclass[twocolumn]{aa}  
\usepackage{graphicx}
\usepackage{rotating}
\usepackage{afterpage}
\usepackage{txfonts}
\usepackage[sort&compress]{natbib}
\bibpunct{(}{)}{;}{a}{}{,}

\newcommand{\um}{$\mu$m}
\newcommand{\brgamma}{Br$\gamma$}

\newcommand\nodata{ ~$\cdots$~ }

\def\degr{\hbox{$^\circ$}}
\def\arcmin{\hbox{$^\prime$}}
\def\arcsec{\hbox{$^{\prime\prime}$}}
\def\utw{\smash{\rlap{\lower5pt\hbox{$\sim$}}}}
\def\udtw{\smash{\rlap{\lower6pt\hbox{$\approx$}}}}

\def\Lsun{\hbox{\it L$_\odot$}}

\def\Msun{\hbox{\it M$_\odot$}}

\def\Teff{\hbox{\it T$_{\rm eff}$}}

\def\Mk{\hbox{\it M$_{\rm K_{\rm s}}$}}
\newcommand{\Ks}{{\it K$_{\rm s}$}}

\newcommand{\Aks}{{\it A$_{\rm K_{\rm s}}$}}
\newcommand{\Ak}{{\it A$_{\rm K}$}}

\def\BCK{\hbox{\it BC$_{\rm K}$}}

\def\simgr{\mathrel{\hbox{\rlap{\hbox{\lower4pt\hbox{$\sim$}}}\hbox{$>$}}}}
\def\Mbol{\hbox{\it M$_{bol}$}}
\def\Mbolone{\hbox{\it M$_{bol1}$}}
\def\Mboltwo{\hbox{\it M$_{bol2}$}}

\begin{document}

\renewcommand{\arraystretch}{0.65}

   \title{Near-infrared spectroscopy of candidate red supergiant stars in  clusters.
      \thanks{Based on observations collected at the European Southern Observatory 
   (ESO Programme 60.A-9700(E), and 089.D-0876, and on observations collected at the UKIRT telescope
   (programme ID H243NS).} 
      \thanks{MM is currently employed by the MPIfR. Part of this work
      was performed at RIT (2009), at ESA (2010), and at the MPIfR.}
      }

\author{Maria Messineo  
	\inst{1,2,3} 
	\and
	Zhu Qingfeng      
	\inst{4} 
 	\and  
	Valentin~D. Ivanov   
	\inst{5} 
	\and 
	Donald~F.  Figer  
	\inst{3} 
	\and
	Ben Davies        
	\inst{6} 
	\and  
	Karl~M.   Menten  
	\inst{1}
	\and  
    Rolf P. Kudritzki 
	\inst{7} 
	\and  
    C.-H. Rosie Chen
	\inst{1} 
}

\institute{ Max-Planck-Institut f\"ur Radioastronomie,
Auf dem H\"ugel 69, D-53121 Bonn, Germany
\email{messineo@mpifr-bonn.mpg.de}
\and
European Space Agency (ESA), The Astrophysics and Fundamental Physics Missions Division, 
Research and Scientific Support Department, Directorate of Science and Robotic Exploration, 
ESTEC, Postbus 299, 2200 AG Noordwijk, The Netherlands
\and
Chester F. Carlson Center for Imaging Science, Rochester Institute
of Technology, 54 Lomb Memorial Drive, Rochester, NY 14623-5604, United States
\and
Key Laboratory for Researches in Galaxies and Cosmology, University of Science and Technology of China, 
Chinese Academy of Sciences, Hefei, Anhui, 230026, China
\and
European Southern Observatory, Alonso de Cordoba 3107, Santiago, Chile 
\and
Astrophysics Research Institute, Liverpool John Moores University,
Twelve Quays House, Egerton Wharf, Birkenhead, Wirral.
CH41 1LD, United Kingdom.    	   
\and
Institute for Astronomy, University of Hawaii, 2680 Woodlawn Drive, Honolulu, HI 96822
}

   \date{Received September 15, 1996; accepted March 16, 1997}

% \abstract{}{}{}{}{} 
% 5 {} token are mandatory
 
  \abstract
   {Clear identifications of Galactic young stellar clusters   farther than a few kpc from the Sun
   are rare,   despite the large number of candidate clusters.
 }
   { We aim to  improve the selection of  candidate clusters rich in massive stars
with a   multiwavelength analysis of photometric  Galactic data that range from optical 
   to mid-infrared wavelengths.
   }
   {We present  a photometric and spectroscopic analysis of five candidate stellar clusters, 
  which were selected as overdensities with bright stars (\Ks$<7$ mag)  in  GLIMPSE and 2MASS images.}
   {A total of 48 infrared  spectra were obtained. 
   The combination of photometry and spectroscopy yielded six new 
   red supergiant stars with masses from 10 M$_\odot$ to 15 M$_\odot$. Two red supergiants are
   located   at Galactic coordinates (l,b)=(16$\rlap{.}^{\circ}$7,$-0\rlap{.}^{\circ}63$)
   and at a  distance of about $\sim 3.9$ kpc; four other red supergiants are 
   members of a cluster at 
   Galactic coordinates (l,b)=(49$\rlap{.}^{\circ}$3,$+0\rlap{.}^{\circ}72$)
   and  at a distance of $\sim 7.0$ kpc.   }
  { 
   Spectroscopic analysis of the brightest  stars of detected overdensities 
   and studies of interstellar extinction 
   along their line of sights are fundamental to distinguish regions of low extinction  from actual stellar clusters.
   The census of young star clusters containing red supergiants is incomplete; 
   in the existing all-sky near-infrared surveys, they  can be identified  as overdensities
   of bright stars with infrared color-magnitude diagrams characterized by gaps.}


   \maketitle
%
%________________________________________________________________

\section{Introduction}

\begin{table*}
\caption{\label{table.targets} List of candidate cluster positions, apparent radii, $<$\Aks$>$, and modulus distances (DM). }
\begin{tabular} {@{\extracolsep{-.09in}}llrrllllll}
\hline
\hline
ID      &  RA(J2000)$^a$        & DEC(J2000)$^a$  & radius      & $<{\it A}_{\it K{\rm s} ~field}>^b$ &  DM($<{\it A}_{\it K_{\rm s} ~field}>)^c$ & $<{\it A}_{\it K{\rm s} ~cRSGs}>$$^d$ & DM($<{\it A}_{\it K_{\rm s}~RSGs}>$)$^e$ & DM(red clump)$^f$ & DM(Sp)$^g$ \\
        & ${\rm [hh ~mm ~ss]}$      &  [$^{\circ}$ $^{\prime}$ $^{\prime\prime}$]    & [$^{\prime\prime}$] & $\rm [mag]$  & $\rm [mag]$ & $\rm [mag]$& $\rm [mag]$ & $\rm [mag]$ & $\rm [mag]$\\
\hline
     cl1.5 &17 53 45.09 &-28 11 08.29 &  73 & $0.45\pm0.08$ & $(12.7\pm0.2)$ &\nodata      &\nodata        & $14.54\pm0.16$   &\nodata \\ %;giant
     cl9.5 &18 10 54.34 &-21 12 37.67 &  60 & $0.70\pm0.11$ & $(13.3\pm0.2)$ & \nodata     &\nodata        & $14.05\pm0.12$   &\nodata \\ %;giant
     cl16.7&18 23 34.40 &-14 39 10.86 &  76 & $0.44\pm0.27$ & $12.7\pm0.9$ & $0.49\pm0.03$ &$12.8 \pm 0.1$ & $12.93\pm0.13$   &\nodata \\ %;rsg     3.97 giant
     cl49.3&19 19 31.53 & 14 54 08.57 & 130 & $0.48\pm0.35$ & $13.8\pm1.9$ & $0.47\pm0.11$ &$13.7 \pm 0.5$ & $14.22\pm0.12$   & $14.16\pm1.00$  \\ %;rsg     3.44 giant
     cl59.8&19 46 18.57 & 23 22 30.23 &  85 & $0.26\pm0.17$ & $12.9\pm1.4$ & \nodata       &\nodata        &  \nodata         &\nodata \\ %;giant
\hline
\end{tabular}
\begin{list}{}{}
\item[] 
{\bf Notes:} $(a)$ Positions are derived as the centroid of  surface brightness maps from  2MASS \Ks\ band images.~
$(b)$ $<{\it A}_{\it K_{\rm s} ~field}>$ is the  average \Aks\ of the spectroscopically observed late-type  stars in the field.~
$(c)$  DM($<{\it A}_{\it K_{\rm s} ~field}>$) is the distance modulus obtained from the \Aks\ values
of  detected late-type stars and the model of \citet{drimmel03}.~
$(d)$ $<{\it A}_{\it K_{\rm s}~RSGs}>$ is the  average \Aks\  of  the detected RSGs.~
$(e)$ DM($<{\it A}_{\it K_{\rm s}~RSGs}>$)  is the distance modulus obtained from the \Aks\ values
of  cRSGs and the model of \citet{drimmel03}.~
$(f)$ DM(red clump) is the distance modulus obtained from the \Aks\ values
of  cRSGs and field red clump stars.~
$(g)$ DM(Sp): spectro-photometric  distance modulus of star 39, by assuming a giant case (O9.5-B3III).
For a dwarf case,  DM would be  $13.02\pm0.77$ mag.
\end{list}
\end{table*}

Stellar clusters are the best approximation in nature of a simple stellar 
population, they are the clocks of the Universe, and the  building blocks of  galaxies.
More than 90\% of massive stars  are found in young and massive stellar clusters \citep{dewit05}.

In the inner regions of the Galactic disk,  the identification of 
stellar clusters is a difficult task because of their irregular shapes
and density profiles, and because of high and patchy interstellar extinction. 
About 3000  new candidate stellar clusters  have  been identified as 
significant peaks of stellar counts  in the 2MASS and GLIMPSE catalogs,
or with visual inspection of images \citep[e.g.,][]{ivanov10,mercer05,dutra03,froebrich07,
borissova11,borissova14}. 
However, only a few of these candidates have been confirmed with follow-up 
spectroscopic and photometric studies,   mostly  on the 
near-side of the Galactic plane \citep[e.g.,][]{messineo09,davies11,chen13}. 
A multiwavelength  photometric screening/selection of candidate 
massive clusters is mandatory for the best  use of  the  observing facilities, 
since spectroscopic follow-up can only be accessible for a limited number 
of candidate clusters.   Typically,  50\% of the detected overdensities are found to be spurious \citep{froebrich07}.
For example, overdensities may result from regions of low interstellar extinction \citep[e.g.,][]{dutra02}.
An overdensity  can be defined as a cluster of stars if it is made of stars at the same distance
and approximately the same age. Depending on the dynamical status, a cluster can be a bound cluster or an association. 
Typically,  cluster members have  similar interstellar extinction, and may be recognized
on color magnitude diagrams (CMDs) as specific sequences.

A new class of stellar clusters has recently been discovered, the red supergiant 
clusters (RSGCs),  which  are characterized by a  large number  of red supergiants 
(RSGs)  \citep{figer06,davies07,clark09,negueruela10,negueruela11,gonzalez12}.
RSGC1, RSGC2, RSGC3, RSGC4, and RSGC5 contain 14, 26,
$\sim 35$, $>13$, and 7 RSGs, respectively, or  collectively $\sim $15\% of all  
known RSGs in the Galaxy \citep{messineo12}.
These newly discovered RSGCs are  all concentrated between 
longitude $l=25$\degr\  and $l=30$\degr,
i.e., close to the near end of the Galactic bar,
 where the bar appears to meet the Scutum-Crux spiral arm.
However,  their census    is incomplete \citep[e.g.,][]{messineo12};
further searches for RSGCs are needed to investigate Galactic structure.
RSGCs are dominated by RSGs, which are intrinsically bright at infrared 
wavelengths.  Their  near-infrared  CMDs 
are characterized by gaps  of several magnitudes between  the RSGs  
and  the blue supergiant  members. 
 RSGCs are detected as overdensities of solely infrared bright stars, 
so bright (typically \Ks $\la 7.0$ mag at 6 kpc) that 
they  can be detected throughout the Galactic plane;
infrared dimmer main sequence members are hard to identify in the glare 
of  bright RSGs.

In this paper, we  analyze  five candidate clusters rich in   infrared bright stars 
 that show different types of CMDs
by means of a quantitative analysis of  near-infrared spectra of their brightest stars. 
In Sect.\  \ref{sectargets}, we describe the observed candidate stellar clusters;
in  Sect.\ \ref{secdata}, we report on the available infrared spectroscopic and 
photometric data.
Spectral and photometric classifications  are given in Sect.\ \ref{secclassification}.
The cluster properties are discussed in Sect. \ref{secclusterings}.
Finally, our findings are summarized in Sect.\  \ref{secsummary}.

\section{Targeted candidate clusters}
\label{sectargets}

\begin{figure*}[!]
\begin{centering}
\resizebox{0.42\hsize}{!}{\includegraphics[angle=0]{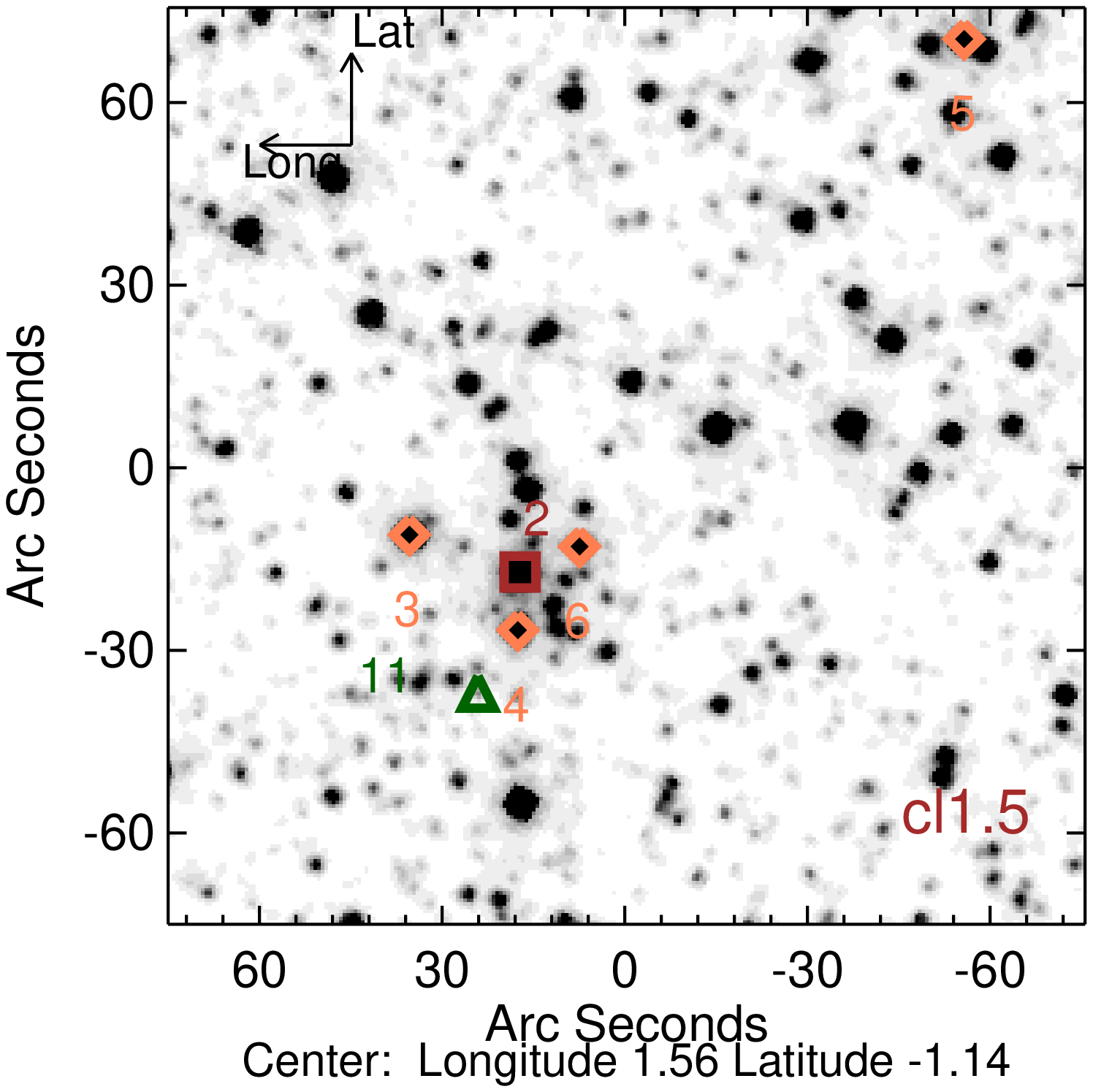}}
\resizebox{0.42\hsize}{!}{\includegraphics[angle=0]{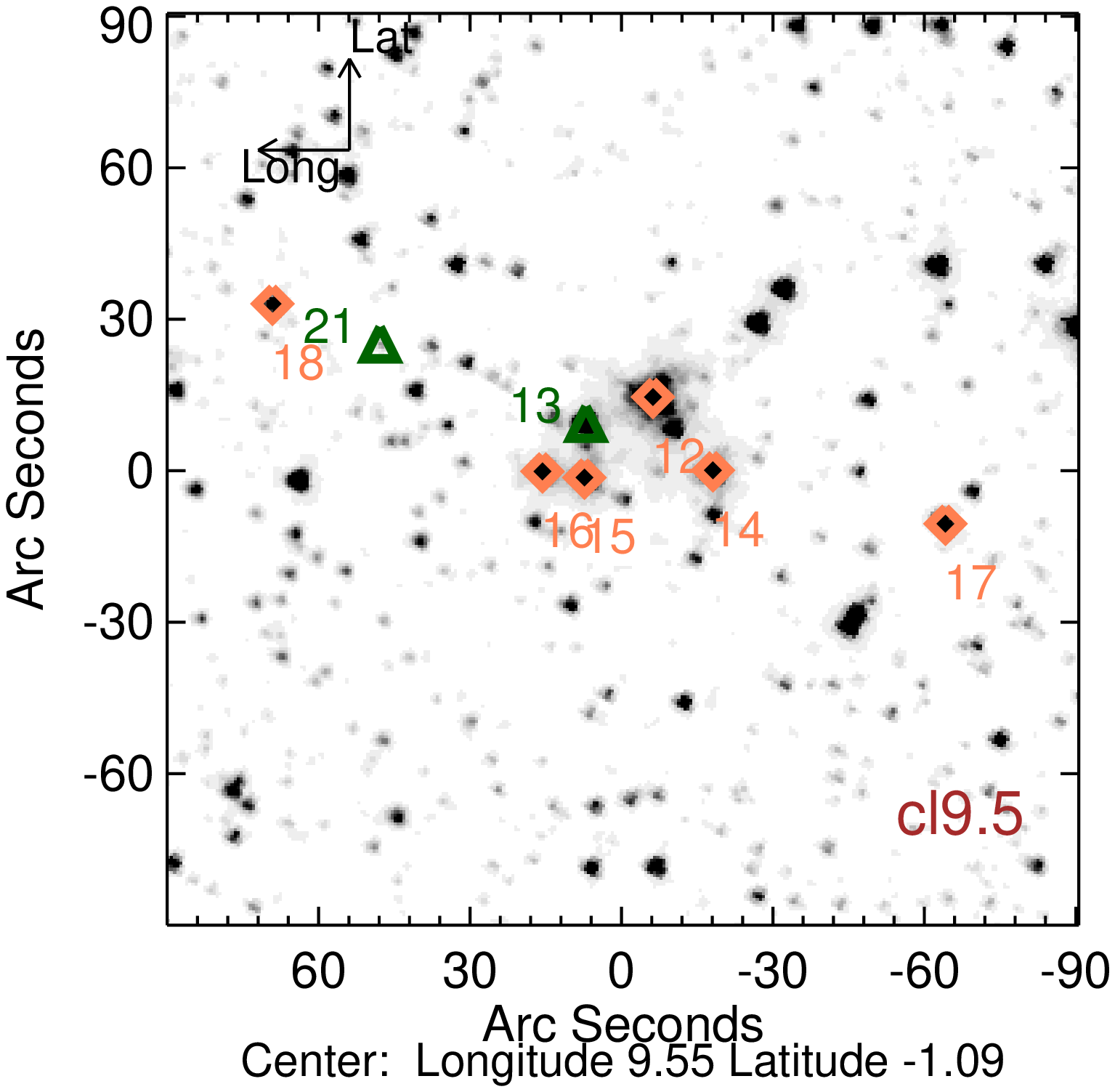}}
\end{centering}
\begin{centering}
\resizebox{0.42\hsize}{!}{\includegraphics[angle=0]{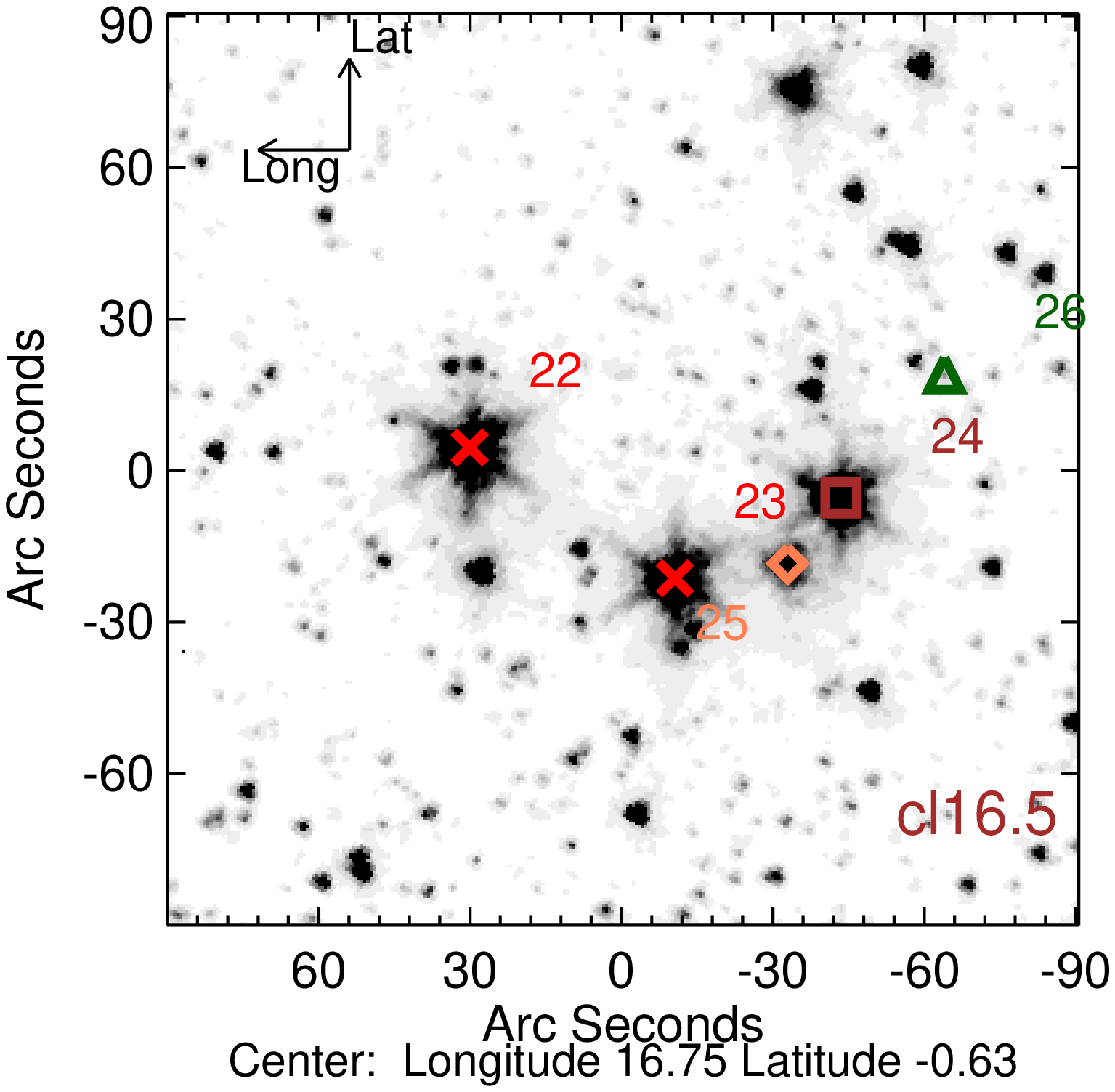}}
\resizebox{0.42\hsize}{!}{\includegraphics[angle=0]{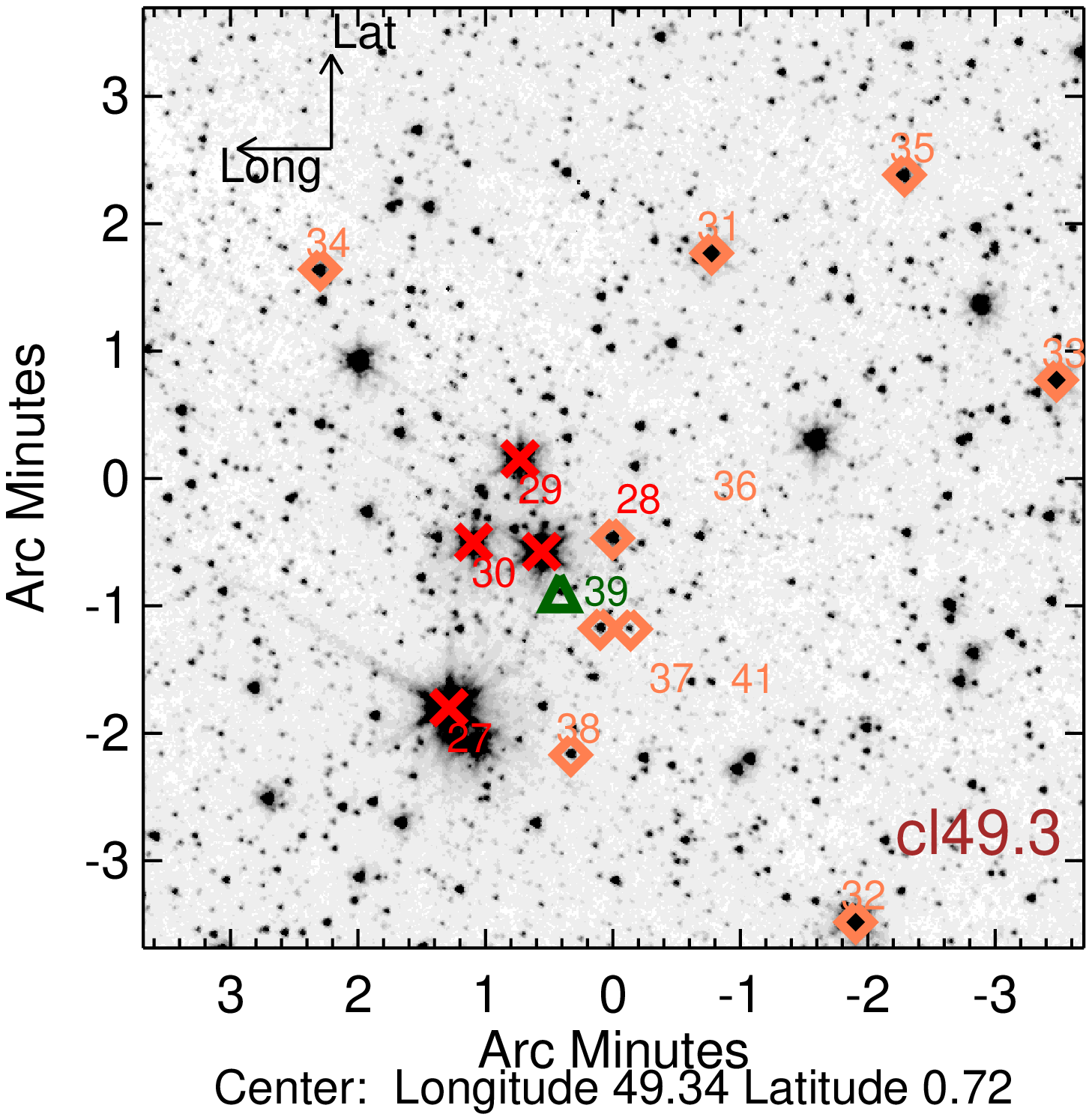}}
\end{centering}
\begin{centering}
\resizebox{0.42\hsize}{!}{\includegraphics[angle=0]{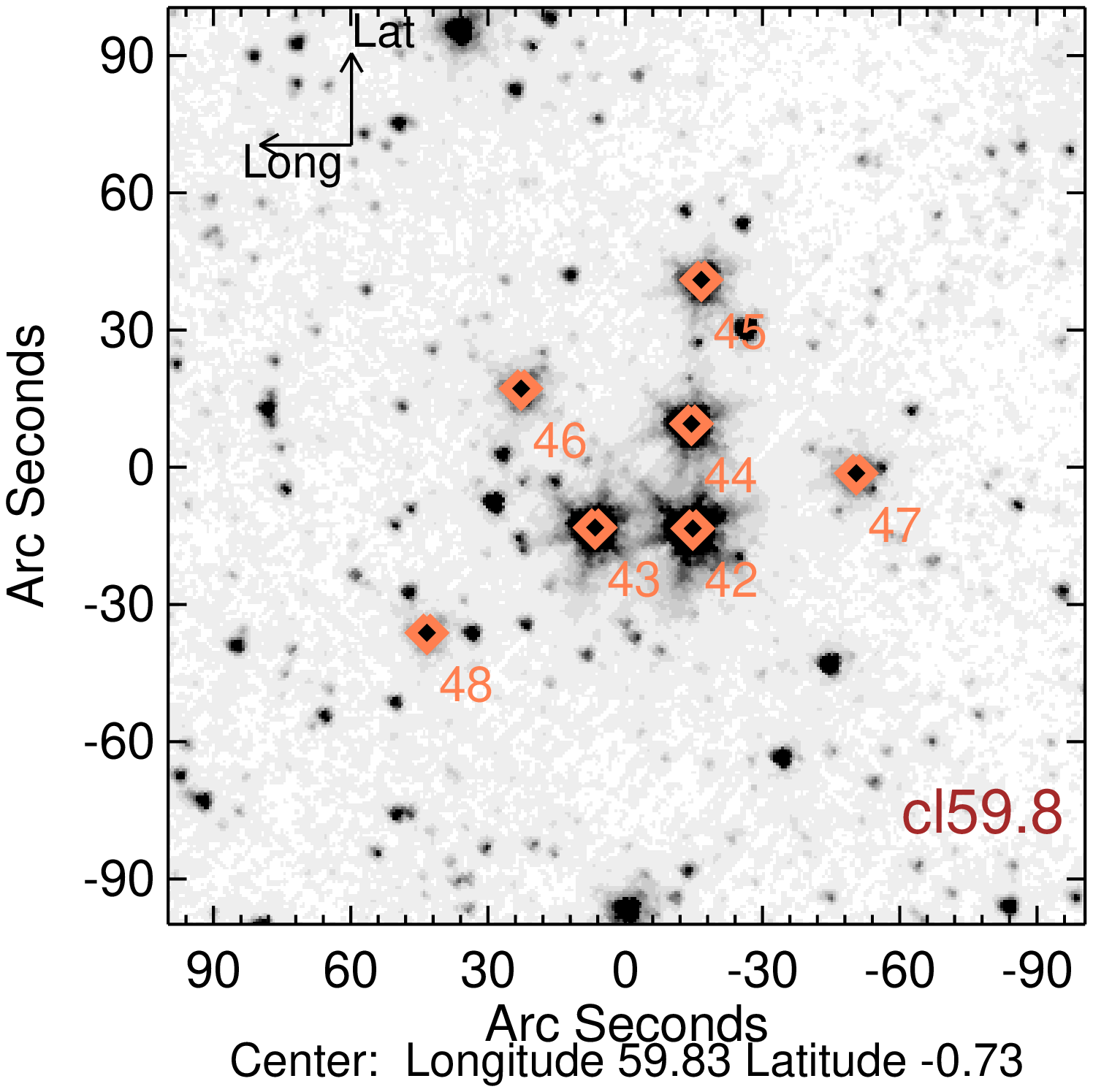}}
\end{centering}
\caption{\label{cl1.5}  GLIMPSE   3.6 \um\  images of the candidate clusters listed
in Table \ref{table.targets}:  cl1.5, cl9.5, cl16.5, cl49.3, and cl59.8. 
Spectroscopic targets are labeled as in 
Tables \ref{table.phot} and \ref{table.spec}.  
Diamonds indicate red giants, squares  Mira-like AGB stars, crosses  RSG stars, 
and triangles  early- and yellow-type stars. 
The coordinates of the image centers  are in degrees.}
\end{figure*}

A random sample of candidate stellar clusters were selected among  overdensities 
of  infrared-bright stars in both GLIMPSE and 2MASS images \citep[see, e.g.,][]{ivanov10,ivanov02}.
The candidate clusters were selected  to have different types of CMDs, 
with gaps and without gaps (see  Sect. \ref{secclusterings}).
The observed candidate clusters are listed in Table \ref{table.targets}.
The list reports their coordinates, which are  
 the flux weighted centroids  of $5$\arcmin $\times5 $\arcmin\ \Ks\ band  images from 2MASS.
The overdensities of  infrared-bright stars are shown in Fig.\ \ref{cl1.5},
and in the star counts    of Table \ref{counts}.

\begin{table}
\caption{\label{counts} Counts of stars with \Ks$> 10.0$, 8.5, and 7.0 mag in the  fields  listed
in Table \ref{table.targets}, and, as a comparison,  
in control fields of equal area.}
\begin{tabular}{@{\extracolsep{-.03in}}l rrr rrr ll}
\hline
\hline
ID &  \multicolumn{3}{c}{Center/Control fields}  \\
\hline
      &  Nstar                & Nstar                 & Nstar                  \\  
      &( $K_{\rm s} <10 $ mag)&($K_{\rm s} <8.5 $ mag)& ($K_{\rm s} < 7 $ mag) \\
\hline
cl1.5 &  57/24 & 8/1 &      1/1&  \\
cl9.5 &  21/06 & 4/1 &      1/1&  \\
cl16.7&  15/09 & 7/2 &   { 4/1}& \\
cl49.3&  27/14 & 7/5 &   { 5/1}& \\
cl59.8&   9/02 & 5/1 &      3/1&  \\
\hline
\end{tabular}
\end{table}

\section{Data}
\label{secdata}

\subsection{Near\--IR spectroscopy}

\begin{sidewaystable*} 
%\vspace*{-3.5cm} %;;;two columns
\vspace*{+16.5cm} %;;;two columns
%\vspace*{+10.5cm}
\caption{\label{table.phot} Infrared counterparts of the spectroscopically observed targets.}  
{\footnotesize
%{\tiny
\begin{tabular}{@{\extracolsep{-.1in}}rrr|rrr|rrr|rrrr|rrrr|rrrr|rrrrr}
\hline 
\hline 
    &  &   &\multicolumn{3}{c}{\rm 2MASS}   &\multicolumn{3}{c}{\rm DENIS} &   \multicolumn{4}{c}{\rm GLIMPSE}   &  \multicolumn{4}{c}{\rm MSX$^a$}& \multicolumn{4}{c}{\rm WISE$^b$} & \multicolumn{3}{c}{\rm NOMAD} &\\ 
\hline 
 {\rm ID}   & {\rm RA(J2000)} & {\rm DEC(J2000)}  &{\it J} & {\it H} & {\it $K_{\rm S}$}  &
 {\it I} & {\it J} & {\it $K_{\rm S}$} & { [3.6]} & { [4.5]} & { [5.8]} & { [8.0]} &
 {\it A} & {\it D} & {\it C}& {\it D} &{\it W1} &{\it W2} & {\it W3} &{\it W4} & {\it B} &{\it V} & {\it R} &\\ 
\hline 
 &{\rm [hh mm ss]}   & {\rm [deg mm ss]}    &{\rm [mag]}   &	{\rm [mag]}    & {\rm [mag]}     & {\rm [mag]} &{\rm [mag]}  & {\rm [mag]}  & 
 {\rm [mag]}  & {\rm [mag]} &{\rm [mag]} &{\rm [mag]}  & {\rm [mag]}&{\rm [mag]}&{\rm [mag]}&{\rm [mag]}&{\rm [mag]}&{\rm [mag]}& 
 {\rm [mag]}& {\rm [mag]}& {\rm [mag]}&{\rm [mag]}&{\rm [mag]}& \\ 
\hline 

                             1  &    17 53 60.00 &   -28 10 50.19 &   7.48 &  5.88 &  4.86 & \nodata & \nodata & \nodata & \nodata & \nodata & \nodata & \nodata &  1.56 &  0.67 &  0.70 &  0.01 & \nodata & \nodata & \nodata & \nodata & \nodata & \nodata & \nodata &  \\
                             2  &    17 53 45.50 &   -28 11 07.12 &   9.68 &  8.17 &  7.35 & \nodata & \nodata & \nodata &  6.94 &  7.00 &  6.69 &  6.46 &  5.88 & \nodata & \nodata &\nodata&  6.74 &  6.72 &  5.72 &  4.78 & \nodata & \nodata & \nodata &  \\
                             3  &    17 53 45.80 &   -28 10 48.35 &  10.57 &  8.91 &  8.09 & \nodata & \nodata & \nodata &  7.45 &  7.55 &  7.20 &  7.08 & \nodata & \nodata & \nodata & \nodata &  7.69 &  7.41 &  6.73 &  6.11 & \nodata & \nodata & \nodata &  \\
                             4  &    17 53 46.14 &   -28 11 11.64 &  10.57 &  9.20 &  8.54 & \nodata & \nodata & \nodata &  8.18 &  8.28 &  8.06 &  8.01 & \nodata & \nodata & \nodata & \nodata & \nodata & \nodata & \nodata & \nodata &  17.89 & \nodata &  17.41 &  \\
                             5  &    17 53 36.99 &   -28 11 25.62 &  10.84 &  9.43 &  8.81 & \nodata & \nodata & \nodata &  8.52 &  8.72 &  8.51 &  8.48 & \nodata & \nodata & \nodata & \nodata &  7.83 &  7.80 &  8.10 & \nodata & \nodata & \nodata &  18.65 &  \\
                             6  &    17 53 44.85 &   -28 11 13.42 &  11.07 &  9.57 &  9.01 & \nodata & \nodata & \nodata &  8.69 &  8.80 &  8.65 &  8.63 & \nodata & \nodata & \nodata & \nodata & \nodata & \nodata & \nodata & \nodata & \nodata & \nodata &  19.54 &  \\
                             7  &    17 53 58.30 &   -28 10 54.96 &  10.95 &  9.67 &  9.07 & \nodata & \nodata & \nodata &  8.80 &  8.84 &  8.59 &  8.73 & \nodata & \nodata & \nodata & \nodata &  8.86 &  9.18 & \nodata & \nodata & \nodata &  13.15 & \nodata &  \\
                             8  &    17 54 03.84 &   -28 10 46.62 &  \nodata & 10.71 & 10.06 & \nodata & \nodata & \nodata &  9.68 &  9.77 &  9.42 &  9.47 & \nodata & \nodata & \nodata & \nodata &  9.64 &  9.74 & \nodata &  6.40 &  18.81 & \nodata &  23.26 &  \\
                             9  &    17 53 35.93 &   -28 11 25.89 &  11.00 & 10.30 & 10.08 & \nodata & \nodata & \nodata & 10.02 & 10.06 & 10.00 &  9.99 & \nodata & \nodata & \nodata & \nodata & \nodata & \nodata & \nodata & \nodata &  14.39 &  13.28 &  13.24 &  \\
                            10  &    17 54 01.74 &   -28 10 49.68 &  12.15 & 10.78 & 10.23 & \nodata & \nodata & \nodata &  9.89 & 10.04 &  9.53 &  9.67 & \nodata & \nodata & \nodata & \nodata & \nodata & \nodata & \nodata & \nodata &  17.28 &  16.42 & \nodata &  \\
                            11  &    17 53 47.05 &   -28 11 11.16 &  12.48 & 12.28 & 12.24 & \nodata & \nodata & \nodata & 12.10 & 12.18 & 12.16 & 12.15 & \nodata & \nodata & \nodata & \nodata & \nodata & \nodata & \nodata & \nodata &  13.37 &  12.75 &  13.10 &  \\

                            12  &    18 10 53.31 &   -21 12 34.75 &   8.93 &  7.18 &  6.39 & 14.69 &  9.33 &  6.84 &  6.35 &  6.04 &  5.75 &  5.54 &  5.15 & \nodata & \nodata &\nodata&  5.54 &  5.80 &  4.89 &  3.81 & \nodata & \nodata & \nodata &  \\
                            13  &    18 10 54.09 &   -21 12 25.57 &   7.99 &  7.96 &  7.83 &  9.03 &  8.52 &  8.28 &  7.78 &  7.79 &  7.74 &  7.80 & \nodata & \nodata & \nodata & \nodata &  7.72 &  7.71 &  7.63 &  7.82 &   9.01 &   8.79 &   8.66 &  \\
                            14  &    18 10 53.81 &   -21 12 52.15 &  10.74 &  8.90 &  8.02 & 17.59 & 11.14 &  8.54 &  7.48 &  7.61 &  7.31 &  7.27 & \nodata & \nodata & \nodata & \nodata &  7.56 &  7.48 &  6.87 &  5.65 & \nodata & \nodata & \nodata &  \\
                            15  &    18 10 54.78 &   -21 12 30.55 &  10.91 &  9.05 &  8.13 & \nodata & 10.80 &  8.05 &  7.51 &  7.79 &  7.41 &  7.31 & \nodata & \nodata & \nodata & \nodata &  7.74 &  7.71 &  6.96 &  6.00 & \nodata & \nodata & \nodata &  \\
                            16  &    18 10 54.99 &   -21 12 22.67 &  11.44 &  9.63 &  8.84 & 16.16 & 11.72 &  9.23 &  8.20 &  8.12 &  7.91 &  7.79 & \nodata & \nodata & \nodata & \nodata &  8.40 &  8.19 &  7.89 &  7.51 & \nodata & \nodata & \nodata &  \\
                            17  &    18 10 52.89 &   -21 13 37.62 &  11.96 & 10.34 &  9.49 & 16.77 & 12.43 &  9.79 & \nodata & \nodata & \nodata & \nodata & \nodata & \nodata & \nodata & \nodata & \nodata & \nodata & \nodata & \nodata & \nodata & \nodata & \nodata &  \\
                            18  &    18 10 54.75 &   -21 11 19.81 &  11.73 & 10.18 &  9.60 & 15.89 & 12.01 &  9.88 &  9.30 &  9.40 &  9.13 &  9.15 & \nodata & \nodata & \nodata & \nodata &  9.19 &  9.27 &  8.69 &  7.14 & \nodata & \nodata &  18.08 &  \\
                            19  &    18 10 54.82 &   -21 10 49.14 &  12.05 & 11.88 & 11.76 & 12.92 & 12.63 & 12.31 & 11.66 & 11.51 & 11.65 & 12.02 & \nodata & \nodata & \nodata & \nodata & \nodata & \nodata & \nodata & \nodata &  12.97 &  12.87 &  11.95 &  \\
                            20  &    18 10 55.33 &   -21 10 15.97 &  12.38 & 12.02 & 11.77 & 13.41 & 13.06 & 12.02 & 11.55 & 11.50 & 11.58 & 11.59 & \nodata & \nodata & \nodata & \nodata & \nodata & \nodata & \nodata & \nodata &  13.90 &  13.64 &  13.33 &  \\
                            21  &    18 10 54.50 &   -21 11 42.23 &  14.22 & 12.96 & 12.49 & \nodata & 14.83 & 13.06 & 12.11 & 12.11 & 11.88 & 12.57 & \nodata & \nodata & \nodata & \nodata & \nodata & \nodata & \nodata & \nodata & \nodata & \nodata & \nodata &  \\

                            22  &    18 23 35.07 &   -14 38 40.87 &   6.38 &  5.15 &  4.55 &  9.41 &  6.22 &  3.97 & \nodata & \nodata &  4.21 &  4.23 &  4.07 & \nodata & \nodata &\nodata&  4.27 &  4.15 &  4.21 &  4.00 &  15.40 &  13.70 &  12.21 &  \\
                            23  &    18 23 35.33 &   -14 39 28.95 &   7.01 &  5.68 &  5.06 & 11.25 &  6.78 &  4.22 &  4.65 &  4.75 &  4.35 &  4.18 &  4.02 &  3.18 &  3.02 &\nodata&  4.75 &  4.14 &  3.66 &  2.60 &  18.98 &  16.47 &  15.04 &  \\
                            24  &    18 23 33.30 &   -14 39 50.57 &   9.74 &  7.34 &  5.91 & 13.77 &  8.93 &  4.80 & \nodata & \nodata &  4.07 &  3.72 &  2.96 &  1.97 &  1.88 &  1.58 &  4.56 &  3.30 &  2.43 &  1.32 &  16.09 &  15.33 &  13.40 &  \\
                            25  &    18 23 34.43 &   -14 39 47.33 &   9.11 &  7.43 &  6.64 & 13.64 &  9.03 &  6.46 &  6.71 &  6.34 &  6.08 &  6.09 & \nodata & \nodata & \nodata & \nodata &  6.15 &  6.41 &  6.45 &  6.45 & \nodata & \nodata & \nodata &  \\
                            26  &    18 23 31.14 &   -14 39 57.17 &  13.12 & 12.52 & 12.10 & 14.55 & 13.00 & 11.89 & 11.86 & 11.92 & 11.60 & \nodata & \nodata & \nodata & \nodata & \nodata & \nodata & \nodata & \nodata & \nodata &  16.20 &  15.36 &  15.00 &  \\
                            
                            27  &    19 19 33.81 &    14 53 55.68 &   5.49 &  4.52 &  4.12 & \nodata & \nodata & \nodata &  3.83 &  3.91 &  3.60 &  3.37 &  3.10 &  2.61 &  2.67 &  1.41 &  3.69 &  3.39 &  2.90 &  2.06 &  14.55 &  17.18 &  10.84 &  \\
                            28  &    19 19 27.91 &    14 53 51.64 &   7.40 &  5.98 &  5.41 & \nodata & \nodata & \nodata &  4.89 &  5.31 &  4.92 &  4.72 &  4.57 & \nodata & \nodata &\nodata&  5.04 &  4.93 &  4.53 &  3.76 & \nodata &  15.00 & \nodata &  \\
                            29  &    19 19 25.58 &    14 54 21.07 &   7.88 &  6.54 &  5.97 & \nodata & \nodata & \nodata & \nodata &  5.91 &  5.57 &  5.46 &  5.49 & \nodata & \nodata &\nodata&  5.62 &  5.69 &  5.44 &  4.83 &  16.92 &  15.09 &  12.95 &  \\
                            30  &    19 19 28.68 &    14 54 21.98 &   8.41 &  6.93 &  6.38 & \nodata & \nodata & \nodata &  5.72 &  6.23 &  5.96 &  5.93 & \nodata & \nodata & \nodata & \nodata &  5.99 &  6.11 &  5.97 &  5.58 &  16.67 &  15.07 &  13.25 &  \\
                            31  &    19 19 16.76 &    14 53 46.72 &   8.35 &  7.49 &  7.16 & \nodata & \nodata & \nodata &  6.97 &  7.16 &  7.00 &  6.97 & \nodata & \nodata & \nodata & \nodata &  6.93 &  7.08 &  7.06 &  7.31 &  14.17 &  12.82 &  10.74 &  \\
                            32  &    19 19 33.77 &    14 50 19.19 &  10.18 &  8.48 &  7.40 & \nodata & \nodata & \nodata &  6.12 &  5.99 &  5.63 &  5.29 &  5.22 & \nodata & \nodata &\nodata&  6.31 &  5.76 &  4.62 &  3.79 &  13.72 &  13.08 &  12.01 &  \\
                            33  &    19 19 15.15 &    14 50 55.29 &  11.09 &  9.11 &  7.93 & \nodata & \nodata & \nodata &  6.79 &  6.65 &  6.22 &  5.89 &  5.46 & \nodata & \nodata &\nodata&  6.52 &  5.98 &  4.65 &  3.74 & \nodata& \nodata& \nodata&  \\
                            34  &    19 19 23.18 &    14 56 26.01 &   9.14 &  8.35 &  8.06 & \nodata & \nodata & \nodata &  7.93 &  8.10 &  7.95 &  7.87 & \nodata & \nodata & \nodata & \nodata &  7.61 &  7.84 &  7.89 &  8.64 &  14.22 &  13.18 &  10.74 &  \\
                            35  &    19 19 11.58 &    14 52 43.78 &   8.86 &  8.28 &  8.09 & \nodata & \nodata & \nodata &  8.02 &  8.08 &  7.96 &  7.96 & \nodata & \nodata & \nodata & \nodata &  7.97 &  8.02 &  7.98 &  8.45 &  12.51 &  11.52 &  10.92 &  \\
                            36  &    19 19 26.44 &    14 53 25.01 &  10.19 &  8.86 &  8.39 & \nodata & \nodata & \nodata &  8.06 &  8.23 &  8.04 &  8.01 & \nodata & \nodata & \nodata & \nodata &  8.05 &  8.17 &  8.28 & \nodata & \nodata& \nodata& \nodata&  \\
                            37  &    19 19 29.23 &    14 53 10.32 &  12.00 & 10.60 & 10.07 & \nodata & \nodata & \nodata &  9.74 &  9.88 &  9.69 &  9.65 & \nodata & \nodata & \nodata & \nodata &  9.76 &  9.86 & 10.38 &  8.52 & \nodata & \nodata & \nodata &  \\
                            38  &    19 19 33.30 &    14 52 54.58 &  12.76 & 11.01 & 10.24 & \nodata & \nodata & \nodata &  9.87 & 10.00 &  9.74 &  9.74 & \nodata & \nodata & \nodata & \nodata &  9.91 &  9.99 & \nodata & \nodata & \nodata & \nodata &  20.68 &  \\
                            39  &    19 19 28.80 &    14 53 35.01 &  12.10 & 11.50 & 11.19 & \nodata & \nodata & \nodata & 10.94 & 10.97 & 10.79 & 10.93 & \nodata & \nodata & \nodata & \nodata & \nodata & \nodata & \nodata & \nodata &  17.47 & \nodata &  13.82 &  \\
                            40  &    19 19 15.15 &    14 49 54.25 &  13.55 & 11.99 & 11.29 & \nodata & \nodata & \nodata & \nodata & \nodata & \nodata & \nodata & \nodata & \nodata & \nodata & \nodata & \nodata & \nodata & \nodata & \nodata & \nodata &  17.92 & \nodata &  \\
                            41$^c$  &    19 19 28.78 &    14 52 57.50 &  12.56 & 11.76 & 11.48 & \nodata & \nodata & \nodata & \nodata & \nodata & \nodata & \nodata & \nodata & \nodata & \nodata & \nodata & \nodata & \nodata & \nodata & \nodata &  17.22 & \nodata &  14.07 &  \\

                            42  &    19 46 18.93 &    23 22 10.72 &   8.95 &  7.32 &  6.49 & \nodata & \nodata & \nodata &  6.01 &  6.01 &  5.68 &  5.49 &  5.25 & \nodata & \nodata &\nodata&  6.02 &  5.82 &  5.07 &  4.14 & \nodata & \nodata &  19.86 &  \\
                            43  &    19 46 19.70 &    23 22 29.28 &   8.58 &  7.29 &  6.73 & \nodata & \nodata & \nodata &  6.39 &  6.67 &  6.40 &  6.37 & \nodata & \nodata & \nodata & \nodata &  6.43 &  6.57 &  6.41 &  6.18 &  18.16 &  16.02 &  14.31 &  \\
                            44  &    19 46 17.50 &    23 22 22.46 &   8.05 &  7.24 &  6.97 & \nodata & \nodata & \nodata &  6.85 &  7.01 &  6.90 &  6.82 & \nodata & \nodata & \nodata & \nodata &  6.80 &  6.94 &  6.86 &  6.54 &  14.46 &  11.40 &  11.58 &  \\
                            45  &    19 46 15.45 &    23 22 36.41 &  11.10 &  9.13 &  8.22 & \nodata & \nodata & \nodata &  7.47 &  7.51 &  7.17 &  7.03 &  6.74 & \nodata & \nodata &\nodata&  7.47 &  7.21 &  6.48 &  5.40 & \nodata & \nodata & \nodata &  \\
                            46  &    19 46 18.38 &    23 22 58.52 &  10.24 &  8.86 &  8.39 & \nodata & \nodata & \nodata &  8.14 &  8.30 &  8.09 &  8.03 & \nodata & \nodata & \nodata & \nodata &  8.13 &  8.20 &  8.06 &  7.27 &  18.44 &  16.72 &  15.43 &  \\
                            47  &    19 46 16.87 &    23 21 45.87 &   9.63 &  8.75 &  8.54 & \nodata & \nodata & \nodata &  8.44 &  8.52 &  8.46 &  8.42 & \nodata & \nodata & \nodata & \nodata &  8.37 &  8.45 &  8.46 &  7.90 &  14.24 &  12.96 &  12.04 &  \\
                            48  &    19 46 22.48 &    23 22 49.50 &  10.67 &  9.45 &  9.03 & \nodata & \nodata & \nodata &  8.80 &  8.97 &  8.76 &  8.74 & \nodata & \nodata & \nodata & \nodata &  8.73 &  8.80 &  8.74 &  7.57 &  18.29 &  16.65 &  15.32 &  \\

\hline
\end{tabular}
}
\begin{list}{}{}
  \item[{\bf Notes:}] ($^{\mathrm{a}}$) Upper-limit magnitudes were removed.~
  ($^{\mathrm{b}}$) Measurements with signal-to-noise ratio smaller than 2 were removed.~
  ($^{\mathrm{c}}$) For star 41, the adopted $J$, $H$, and $K$ measurements are from the UKIDSS survey.~
\end{list}
\end{sidewaystable*}

\subsubsection{UIST spectra} 

A set of spectroscopic data was taken with the UKIRT 1-5 micron Imager Spectrometer  
\citep[UIST;][]{ramsay98} on Mauna Kea under program ID H243NS (PI: Kudritzki) on 
2008 July 24. We used  the long-K  grism, which  covers from 2.204 \um\ to 2.513 \um\ 
at a resolving power (R) of 1900. 
Integration times  varied from 10 s to 45 s per exposure, and the number of exposures 
varied from 8  to 20.  

\begin{figure}[!]
\begin{centering}
\resizebox{1.0\hsize}{!}{\includegraphics[angle=0]{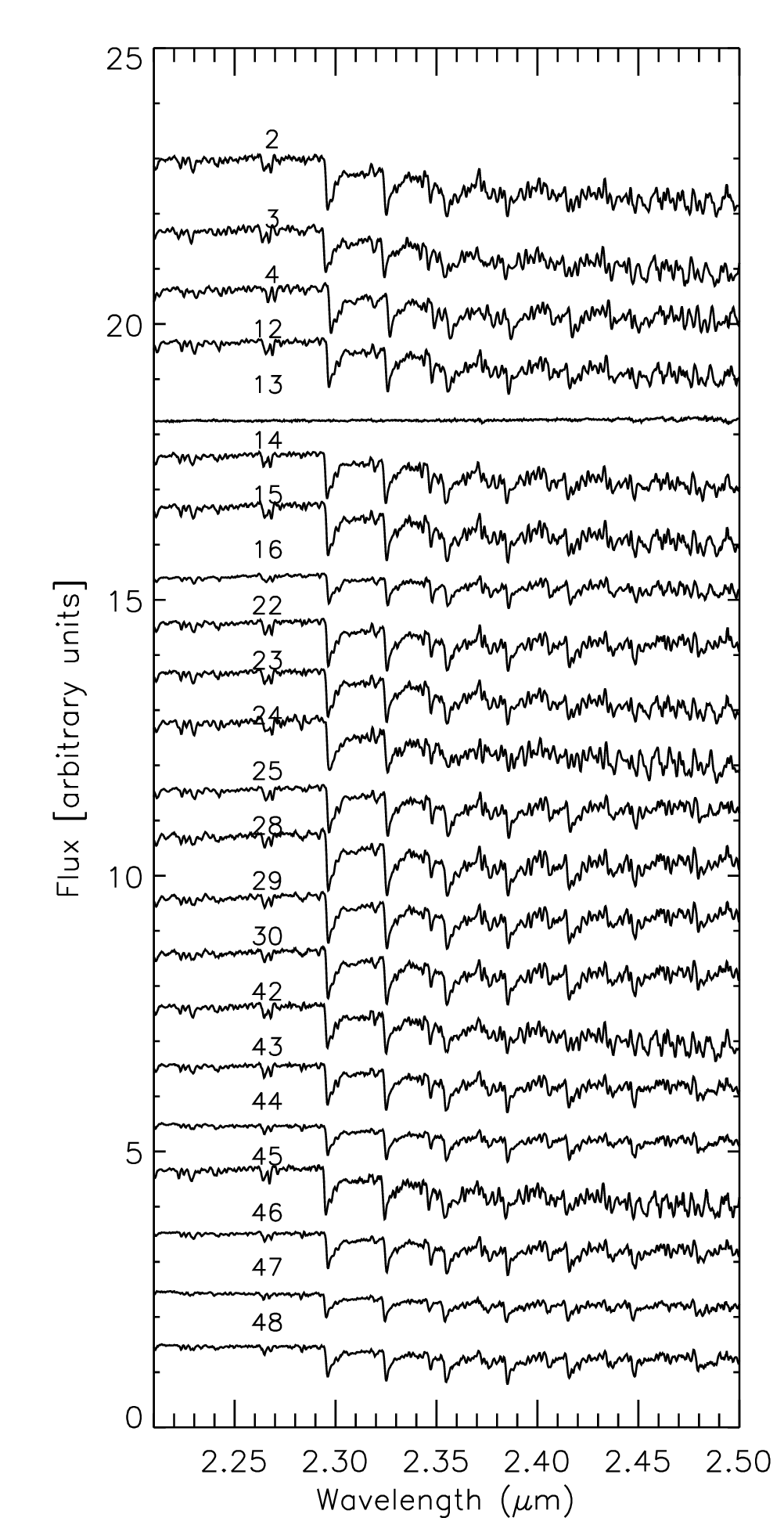}} %;;2columns
%\resizebox{0.7\hsize}{!}{\includegraphics[angle=0]{fig2.eps}}
\end{centering}
\caption{\label{uist.spectra} Long-$K$ spectra taken with UIST.
Identification numbers refer to Table 
\ref{table.spec}. All are late-type stars, with the exception of star 13.}
\end{figure}

Pairs of adjacent  frames at different nod-positions  were subtracted from each other 
and flat-fielded,  wavelength-calibrated  with  Ar arc lines, and corrected 
for atmospheric absorption and instrumental response. 
Curves of atmospheric absorption and instrumental response
were generated by dividing the observed 
spectra  of standard stars (with spectral types from B2 to B9) by blackbody curves.
Linear interpolation was used to remove
\brgamma\ lines and possible \ion{He}{I} lines   from the spectra of 
the standards.
A  total of  22 stars were observed with UKIRT, and are listed in Tables 
\ref{table.phot} and \ref{table.spec}; their spectra are shown in Fig.\    \ref{uist.spectra}.

\subsubsection{SOFI spectra}

For four of the targeted candidate clusters, additional  low-resolution and 
medium-resolution spectra were obtained with the SofI 
spectrograph mounted on the NTT telescope \citep{moorwood98}. Data were taken 
under  program 60.A-9700(E)  at Paranal-La Silla Observatory on 2010 August 3,
and under program 089.D-0876 on 2012 June 1.
The pixel scale is 0\farcs288 pix$^{-1}$.  
 The low-resolution  Red grism  in combination with the   $0\farcs6 \times 290\arcsec$ slit
yielded  R $\approx$980  over the range $\sim$1.50$~ \mu$m to $\sim$2.49$~ \mu$m. 
A few   spectra were taken with the HR grism, the \Ks\ 
filter, and a $1\farcs0$-wide slit  (R $\approx 1900$).

Typically, two slit positions per cluster were observed. The rotator angle was set
to optimize the number of  stars falling  onto the slit. For each slit, frames were taken in a 
nodding sequence ABBA, which included a small jittering between positions, with 
detector integration times (DITs) from 1.18 s to 120 s. 
B-type stars  were selected  as standards, and   observed  with the same settings as the targets.

\begin{figure*}[!]
\begin{centering}
\resizebox{0.49\hsize}{!}{\includegraphics[angle=0]{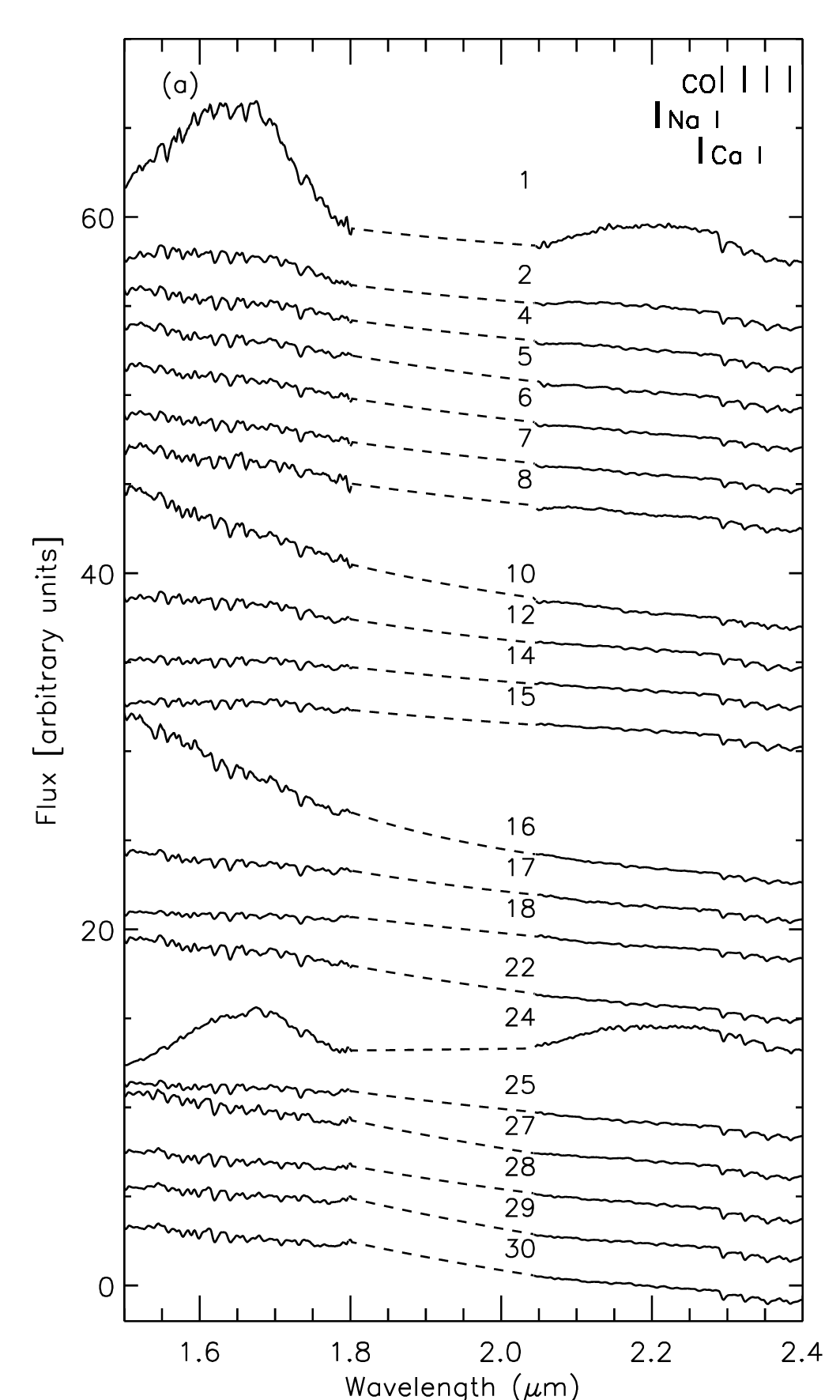}}
\resizebox{0.49\hsize}{!}{\includegraphics[angle=0]{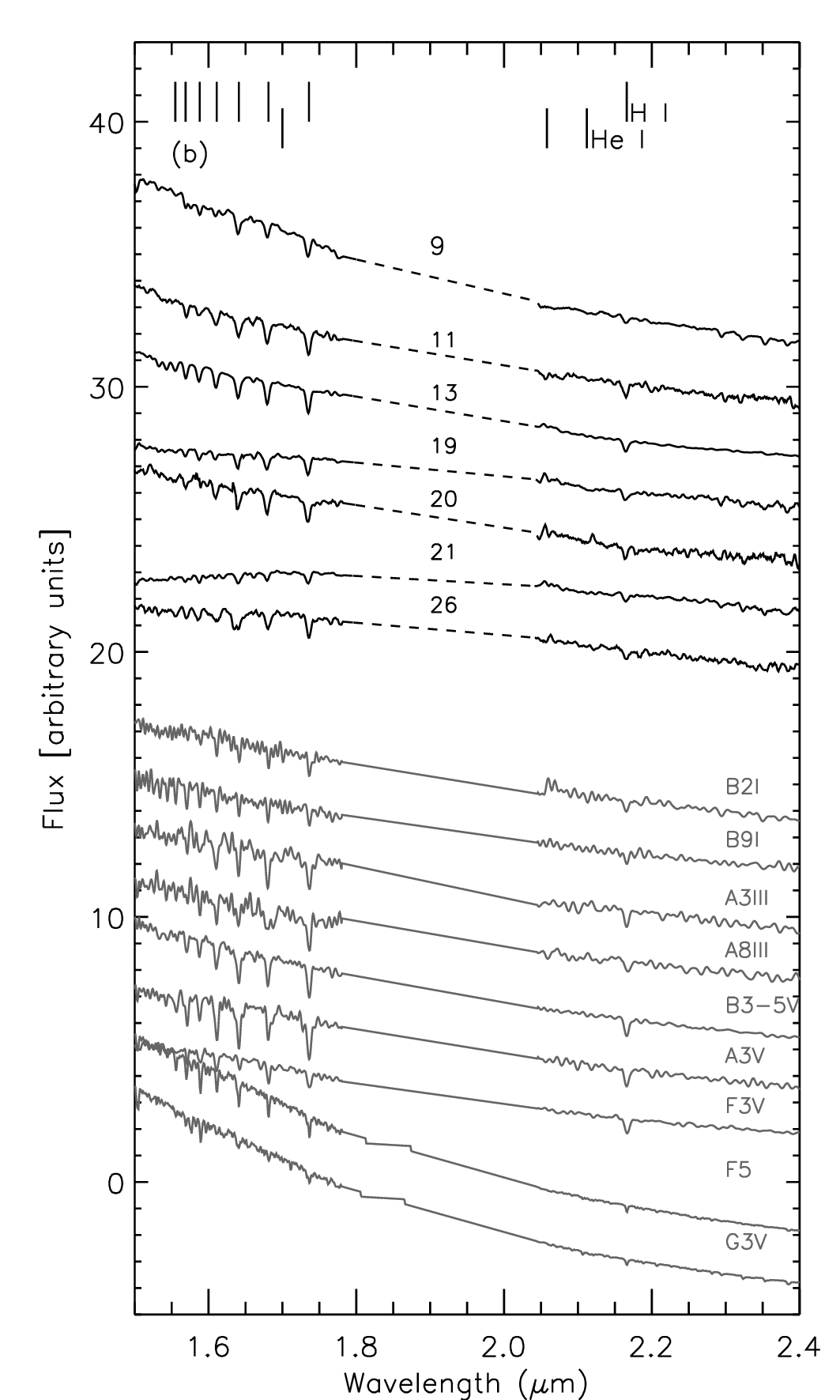}}
\end{centering}
\caption{\label{sofi.spectra} Low-resolution $H$ and $K$ spectra taken with SofI.
Identification numbers are taken from Table \ref{table.spec}. Late-type stars are shown 
in   panel {\it (a)}. 
Early-type stars are shown in panel {\it (b)} together with a few low-resolution template spectra from the
library of \citet{lancon92}, and medium-resolution template spectra from the IRTF library
\citep{rayner09}.
}
\end{figure*}

\begin{figure}[!]
\begin{centering}
\resizebox{0.99\hsize}{!}{\includegraphics[angle=0]{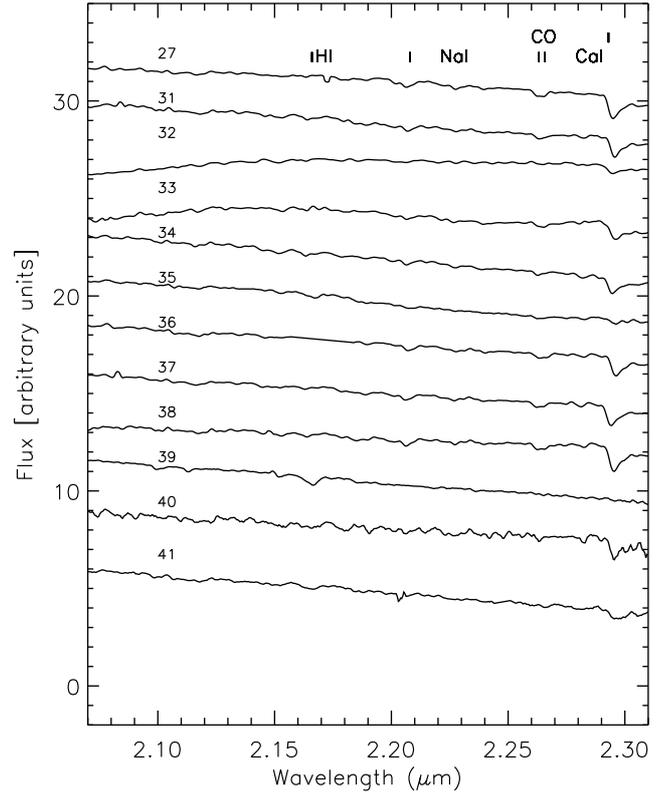}}
\end{centering}
\caption{\label{sofimed.spectra} Medium-resolution  $K$ spectra taken with SofI.
Identification numbers are taken from Table \ref{table.spec}. 
All stars are late-type stars, with the exception of  star 39.
}
\end{figure}

Data reduction was carried out with IDL programs, and  the Image Reduction and Analysis Facility (IRAF),
using the NOAO/onedspec package\footnote{IRAF is distributed by the National Optical Astronomy
Observatories, which is operated by the Association of Universities for Research
in Astronomy, Inc., under cooperative agreement with the National
Science Foundation.}. Pairs of subsequent images  at different nod-positions were subtracted and
flat-fielded. Wavelength calibration was performed with observations of arc lamps.
Only spectra with a signal-to-noise ratio above 30 were used. 
For each  star,  typically,  four spectra were combined and then 
corrected for atmospheric transmission and instrumental response. 
Hydrogen and helium lines were removed from the spectra of the standard stars 
with a linear interpolation; the intrinsic continuum slopes of the standard spectra 
were eliminated by dividing them by 
 blackbody curves (calculated at the \Teff\ of the standard stars).
A total of 28 low-resolution spectra  and 13 medium-resolution spectra were obtained
(see Table \ref{table.spec}, Fig.\ \ref{sofi.spectra}, and Fig.\ \ref{sofimed.spectra}).

\subsection{Photometric data}

For the target stars, simultaneous photometric  measurements in bands $J$,
$H$, and \Ks\  were 
available in  the  Two Micron all Sky Survey (2MASS) catalog \citep{skrutskie06}, with 1\farcs0
resolution; for two fields, additional near-infrared measurements (simultaneous $I$, $J$, and \Ks ) were 
available from the Deep Near-Infrared Survey (DENIS)  catalog \citep{epchtein99}. Deep  $J$, $H$, and $K$  photometry 
was obtained from the UKIDSS Galactic Plane Survey 
\citep[][]{lucas08,hodgkin09}.

Mid-infrared data were available from  the 
Galactic Legacy Infrared Mid--Plane Survey Extraordinaire (GLIMPSE) with the Spitzer Space Telescope, 
from the Midcourse Space Experiment (MSX) Spatial Infrared Imaging Telescope (SPIRIT III), 
and from the Wide-field Infrared Survey Explorer (WISE)
satellite \citep{egan03, benjamin03,fazio04, wright10}. 
The SPITZER/IRAC camera acquired  simultaneous images 
in the four channels at 3.6, 4.5, 5.8, and 8.0 \um, with a  spatial resolution of $1\rlap{.}^{\prime\prime}2$ 
and a sensitivity 
of 2 mJy. MSX  covered from 8 \um\ to 21 \um, with a pixel scale of 18\farcs3, and a 
sensitivity of 0.1 Jy at the short-wavelength.  WISE bands are centered at 3.4, 4.6, 12, 
and 22 \um;   detectability limits are 0.08, 0.11, 1, and  6 mJy; 
spatial resolutions are $6\rlap{.}^{\prime\prime}1$, $6\rlap{.}^{\prime\prime}4$, $6\rlap{.}^{\prime\prime}5$, 
and $12\rlap{.}^{\prime\prime}0$, respectively. 

DENIS and 2MASS data were cross-matched using a search radius of $2\rlap{.}^{\prime\prime}0$; 
2MASS astrometry was retained. 
Mid-infrared counterparts from the MSX catalog were searched within
a radius of 5\arcsec;  matches from the WISE and GLIMPSE catalogs
within  a radius of 2\arcsec.
In addition, we checked for possible visual counterparts in the NOMAD catalog of 
\citet{zacharias05}, using a search radius of 2\arcsec.

Photometric measurements of the spectroscopic targets are listed in Table \ref{table.phot}.

\subsubsection{UKIDSS photometry}
We used JHK photometry from the UKIDSS Galactic Plane Survey (GPS). For every position, JHK
images  were obtained from the UKIDSS archive. Raw frames were processed by the standard
UKIDSS pipeline \citep[][]{lucas08}; corrections for linearity,
dark, flat-fielding,  decurtaining (the removal of a pseudo-periodic ripple),  defringing  (the
removal of interference fringes due to atmospheric emission lines), sky-subtraction, and
cross-talk were applied.

The original WFCAM pixel scale is of 0\farcs4 pix$^{-1}$; however, GPS observations were performed
with a  $2\times 2$ micro-stepping technique, which improved the sampling;
a microstep of $11.5$ pixels was used.
Each observation comprises eight individual exposures of duration 10, 10, and 5 s to make up    
integration times on sources of 80, 80, and 40 s in the $J$, $H$, and $K$ filters, respectively. 
We resampled the exposures of each observation in a finer grid ($3\times3$) in order to 
reduce the image noise, and to improve the detectability of  
faint sources; we used  a bilinear interpolation. Exposures were combined with 
a three $\sigma$ clipping to eliminate 
possible cosmic rays. The astrometric distortion of the final mosaic was corrected by using a set of 
unsaturated stars with 2MASS 
photometry, and performing a polynomial spatial de-warping of third order. 
Source extraction was performed using the DAOPHOT  PSF-fitting algorithm  by \citet{stetson87}.  
For each mosaic, a set of bright and isolated  unsaturated stars were selected, 
and a invariable  PSF was modeled. 
The detection threshold  was set to 4 $\sigma$ (the standard deviation). 
Finally, individual $J$, $H$, and $K$ catalogs were photometrically calibrated 
with overlapping 2MASS sources, positionally cross-correlated, and combined.

For saturated stars that could be uniquely associated with a 2MASS point source,
2MASS magnitudes were retained.

\subsection{Information available on SIMBAD}
The SIMBAD database reports information only for one target.
We classified star 13 as a B0-F0 star; the star coincides 
with  HD166307, which has been  classified as an 
A0III star  at optical wavelength \citep{houk88}.

\section{Stellar classification}
\label{secclassification}

We detected candidate RSGs in two fields.
In order to confirm RSGs in the direction of the inner disk of the Milky Way, a complex procedure is required.
Spectroscopic observations alone cannot firmly distinguish between RSGs and red giant stars.
When analyzing  RSGs, it is crucial to combine  spectroscopic information   
(see Sect.\ \ref{subsec-spectralclass}) with photometric quantities 
(see Sects. \ref{subsec-photometric} and \ref{int.col}),  and to estimate their fundamental  parameters 
(effective temperatures, \Teff,  and luminosities).
Distances of late-type stars are usually estimated in two ways, either  approximated with
kinematic distances, or  estimated using
interstellar extinction as an indicator of distance  \citep[e.g.,][]{habing06,messineo05,drimmel03}. In the  
inner Galaxy kinematic distances alone cannot be trusted as they rely  
on the assumption of circular orbits that  is not valid in the inner 3.5 kpc \citep[e.g.,][]{devaucouleurs64};  
our estimates of distances are based  on interstellar extinction (Sect.\ \ref{secdist}).
Luminosities are described in Sect.\ \ref{subsection-luminosities}.

\begin{sidewaystable*} 
%\vspace*{+5.5cm}
%\vspace*{+18cm}
\vspace*{+16cm}
\caption{\label{table.spec} 
List of stars spectroscopically observed with UKIRT/UIST and NTT/SofI. 
Identification numbers are  as in  Table \ref{table.phot}. 
}  
{\footnotesize
\begin{tabular}{@{\extracolsep{-.04in}}r|rrr|rrr|lll|lll|rl|lr}
\hline 
\hline 
 &\multicolumn{3}{c}{\rm Photometry}   & \multicolumn{12}{c}{\rm Spectroscopy}   \\ 
\hline 
 {\rm ID}   & {\it Q1 }  & {\it Q2}   & {\rm var } & 
\multicolumn{3}{c}{\rm EW(CO)$^a$}  &
\multicolumn{3}{c}{\rm  sp\_RSG$^b$ }&
\multicolumn{3}{c}{\rm   sp\_giant$^c$} &
 {\rm H$_2$O}& Comment &  Instrument \& mode$^d$ &  \\
\hline 
            &            &            &             &UIST  &  SofIl &SofIm & UIST &  SofIl &SofIm &UIST &  SofIl &SofIm & \\ 
            & {\rm [mag]}& {\rm [mag]}&             & [\AA]& [\AA]  &[\AA] &      &        &      &     &        &      & [\%]\\ 
\hline 
 1  &  $-$0.24 $\pm$  0.12  &  \nodata $\pm$ \nodata  &    0  &    \nodata &      53  &   \nodata &      \nodata  &     M1.5  &     \nodata  &      \nodata  &      $>$M7  &     \nodata  &   39  &  sp\_Mira\_AGB  &                      SofIl  \\
 2  &   0.04 $\pm$  0.09  &  $-$0.07 $\pm$  0.09  &    0  &       51  &      52  &   \nodata &      M1.5  &     M1.5  &     \nodata  &       $>$M7  &      $>$M7  &     \nodata  &   15  &  sp\_Mira\_AGB  &                 SofIl, UIST  \\
 3  &   0.20 $\pm$  0.09  &  $-$0.26 $\pm$  0.12  &    0  &       44  &   \nodata &   \nodata &      K5.5  &     \nodata  &     \nodata  &       $>$M7  &     \nodata  &     \nodata  & \nodata &             &                       UIST  \\
 4  &   0.17 $\pm$  0.13  &   0.62 $\pm$  0.13  &    0  &       42  &      46  &   \nodata &        K5  &       M0  &     \nodata  &      M6.5  &      $>$M7  &     \nodata  &    7  &             &                 SofIl, UIST  \\
 5  &   0.30 $\pm$  0.21  &   1.13 $\pm$  0.24  &    0  &    \nodata &      44  &   \nodata &      \nodata  &     K5.5  &     \nodata  &      \nodata  &      $>$M7  &     \nodata  &    7  &             &                      SofIl  \\
 6  &   0.48 $\pm$  0.13  &   1.03 $\pm$  0.14  &    0  &    \nodata &      32  &   \nodata &      \nodata  &       K2  &     \nodata  &      \nodata  &     M1.5  &     \nodata  &    4  &             &                      SofIl  \\
 7  &   0.21 $\pm$  0.10  &   0.95 $\pm$  0.15  &    0  &    \nodata &      36  &   \nodata &      \nodata  &       K3  &     \nodata  &      \nodata  &     M3.5  &     \nodata  &    4  &             &                      SofIl  \\
 8  &  \nodata $\pm$ \nodata  &  \nodata $\pm$ \nodata  &    0  &    \nodata &      38  &   \nodata &      \nodata  &       K4  &     \nodata  &      \nodata  &     M4.5  &     \nodata  &    3  &             &                      SofIl  \\
 9  &   0.30 $\pm$  0.11  &   0.68 $\pm$  0.15  &    0  &    \nodata &   \nodata &   \nodata &      \nodata  &     \nodata  &     \nodata  &      \nodata  &     \nodata  &     \nodata  & \nodata &             &                      SofIl  \\
10  &   0.40 $\pm$  0.11  &   0.40 $\pm$  0.23  &    0  &    \nodata &      33  &   \nodata &      \nodata  &     K2.5  &     \nodata  &      \nodata  &     M2.5  &     \nodata  &    1  &             &                      SofIl  \\
11  &   0.11 $\pm$  0.20  &  $-$0.01 $\pm$  0.51  &    0  &    \nodata &   \nodata &   \nodata &      \nodata  &     \nodata  &     \nodata  &      \nodata  &     \nodata  &     \nodata  & \nodata &             &                      SofIl  \\
12  &   0.32 $\pm$  0.14  &   0.26 $\pm$  0.12  &    0  &       43  &      47  &   \nodata &        K5  &       M0  &     \nodata  &        M7  &      $>$M7  &     \nodata  &    6  &             &                 SofIl, UIST  \\
13  &  $-$0.20 $\pm$  0.18  &   0.07 $\pm$  0.16  &    1  &    \nodata &   \nodata &   \nodata &      \nodata  &     \nodata  &     \nodata  &      \nodata  &     \nodata  &     \nodata  & \nodata &             &                      SofIl, UIST  \\
14  &   0.24 $\pm$  0.08  &   0.71 $\pm$  0.11  &    1  &       42  &      45  &   \nodata &        K5  &       M0  &     \nodata  &        M7  &      $>$M7  &     \nodata  &    2  &             &                 SofIl, UIST  \\
15  &   0.23 $\pm$  0.08  &   0.57 $\pm$  0.11  &    0  &       49  &      45  &   \nodata &        M1  &       M0  &     \nodata  &       $>$M7  &      $>$M7  &     \nodata  &    3  &             &                 SofIl, UIST  \\
16  &   0.40 $\pm$  0.09  &  $-$0.22 $\pm$  0.16  &    0  &       30  &      33  &   \nodata &        K2  &     K2.5  &     \nodata  &      M0.5  &       M2  &     \nodata  &   $-$2  &             &                 SofIl, UIST  \\
17  &   0.08 $\pm$  0.13  &  \nodata $\pm$ \nodata  &    1  &    \nodata &      37  &   \nodata &      \nodata  &     K3.5  &     \nodata  &      \nodata  &       M4  &     \nodata  &   $-$3  &             &                      SofIl  \\
18  &   0.51 $\pm$  0.09  &   0.90 $\pm$  0.14  &    0  &    \nodata &      32  &   \nodata &      \nodata  &     K2.5  &     \nodata  &      \nodata  &     M1.5  &     \nodata  &   $-$4  &             &                      SofIl  \\
19  &  $-$0.04 $\pm$  0.13  &   1.01 $\pm$  0.66  &    1  &    \nodata &   \nodata &   \nodata &      \nodata  &     \nodata  &     \nodata  &      \nodata  &     \nodata  &     \nodata  & \nodata &             &                      SofIl  \\
20  &  $-$0.09 $\pm$  0.16  &   0.13 $\pm$  0.29  &    1  &    \nodata &   \nodata &   \nodata &      \nodata  &     \nodata  &     \nodata  &      \nodata  &     \nodata  &     \nodata  & \nodata &             &                      SofIl  \\
21  &   0.44 $\pm$  0.18  &   1.94 $\pm$  0.97  &    1  &    \nodata &   \nodata &   \nodata &      \nodata  &     \nodata  &     \nodata  &      \nodata  &     \nodata  &     \nodata  & \nodata &             &                      SofIl  \\
22  &   0.16 $\pm$  0.08  &   0.96 $\pm$  0.10  &    0  &       42  &      41  &   \nodata &        K5  &     K4.5  &     \nodata  &        M7  &       M6  &     \nodata  &    4  &             &                 SofIl, UIST  \\
23  &   0.22 $\pm$  0.10  &  $-$0.44 $\pm$  0.11  &    0  &       44  &   \nodata &   \nodata &      K5.5  &     \nodata  &     \nodata  &       $>$M7  &     \nodata  &     \nodata  & \nodata &             &                       UIST  \\
24  &  $-$0.15 $\pm$  0.09  &  $-$2.08 $\pm$  0.12  &    1  &       56  &      29  &   \nodata &        M3  &     K1.5  &     \nodata  &       $>$M7  &       M0  &     \nodata  &   26  &  sp\_Mira\_AGB  &                 SofIl, UIST  \\
25  &   0.28 $\pm$  0.14  &   0.98 $\pm$  0.11  &    0  &       41  &      39  &   \nodata &        K5  &       K4  &     \nodata  &        M6  &       M5  &     \nodata  &   $-$1  &             &                 SofIl, UIST  \\
26  &  $-$0.16 $\pm$  0.21  &  \nodata $\pm$ \nodata  &    0  &    \nodata &   \nodata &   \nodata &      \nodata  &     \nodata  &     \nodata  &      \nodata  &     \nodata  &     \nodata  & \nodata &             &                      SofIl  \\
27  &   0.25 $\pm$  0.70  &  $-$0.66 $\pm$  1.42  &    0  &    \nodata &      39  &      31  &      \nodata  &       K4  &       K2  &      \nodata  &       M5  &     M0.5  &    3  &             &                SofIl, SofIm  \\
28  &   0.39 $\pm$  0.13  &   0.14 $\pm$  0.10  &    0  &       51  &      50  &   \nodata &      M1.5  &       M1  &     \nodata  &       $>$M7  &      $>$M7  &     \nodata  &    1  &             &                 SofIl, UIST  \\
29  &   0.32 $\pm$  0.12  &   0.53 $\pm$  0.10  &    0  &       43  &      47  &   \nodata &      K5.5  &       M0  &     \nodata  &        M7  &      $>$M7  &     \nodata  &    4  &             &                 SofIl, UIST  \\
30  &   0.49 $\pm$  0.11  &   0.84 $\pm$  0.12  &    0  &       42  &      44  &   \nodata &        K5  &     K5.5  &     \nodata  &        M7  &      $>$M7  &     \nodata  &    1  &             &                 SofIl, UIST  \\
31  &   0.28 $\pm$  0.14  &   0.67 $\pm$  0.11  &    0  &     \nodata &   \nodata &      22  &      \nodata  &     \nodata  &       $<$K0 &      \nodata  &     \nodata  &       K3  & \nodata &             &                      SofIm  \\
32  &  $-$0.24 $\pm$  0.14  &  $-$2.90 $\pm$  0.11  &    0  &     \nodata &   \nodata &      28  &      \nodata  &     \nodata  &     K1.5  &      \nodata  &     \nodata  &     K5.5  & \nodata &             &                      SofIm  \\
33  &  $-$0.15 $\pm$  0.12  &  $-$2.31 $\pm$  0.09  &    0  &     \nodata &   \nodata &      27  &      \nodata  &     \nodata  &       K1  &      \nodata  &     \nodata  &       K5  & \nodata &             &                      SofIm  \\
34  &   0.28 $\pm$  0.14  &   0.57 $\pm$  0.13  &    1  &     \nodata &   \nodata &      21  &      \nodata  &     \nodata  &       $<$K0 &      \nodata  &     \nodata  &       K2  & \nodata &             &                      SofIm  \\
35  &   0.23 $\pm$  0.13  &   0.44 $\pm$  0.10  &    0  &     \nodata &   \nodata &       6  &      \nodata  &     \nodata  &       $<$K0 &      \nodata  &     \nodata  &       $<$K0 & \nodata &             &                      SofIm  \\
36  &   0.48 $\pm$  0.10  &   0.80 $\pm$  0.12  &    0  &     \nodata &   \nodata &      24  &      \nodata  &     \nodata  &       $<$K0 &      \nodata  &     \nodata  &     K3.5  & \nodata &             &                      SofIm  \\
37  &   0.46 $\pm$  0.09  &   0.79 $\pm$  0.12  &    0  &     \nodata &   \nodata &      24  &      \nodata  &     \nodata  &     K0.5  &      \nodata  &     \nodata  &     K3.5  & \nodata &             &                      SofIm  \\
38  &   0.36 $\pm$  0.10  &   1.17 $\pm$  0.12  &    0  &     \nodata &   \nodata &      30  &      \nodata  &     \nodata  &       K2  &      \nodata  &     \nodata  &       M0  & \nodata &             &                      SofIm  \\
39  &   0.03 $\pm$  0.10  &   0.19 $\pm$  0.20  &    0  &     \nodata &   \nodata &   \nodata &      \nodata  &     \nodata  &     \nodata  &      \nodata  &     \nodata  &     \nodata  & \nodata &             &                      SofIm  \\
40  &   0.30 $\pm$  0.18  &  \nodata $\pm$ \nodata  &    0  &     \nodata &   \nodata &      22  &      \nodata  &     \nodata  &       $<$K0 &      \nodata  &     \nodata  &       K3  & \nodata &             &                      SofIm  \\
41  &   0.30 $\pm$  2.78  &  \nodata $\pm$ \nodata  &    0  &     \nodata &   \nodata &      15  &      \nodata  &     \nodata  &       $<$K0 &      \nodata  &     \nodata  &       $<$K0 & \nodata &             &                      SofIm  \\
42  &   0.11 $\pm$  0.12  &  $-$0.22 $\pm$  0.11  &    0  &       44  &   \nodata &   \nodata &      K5.5  &     \nodata  &     \nodata  &       $>$M7  &     \nodata  &     \nodata  & \nodata &             &                       UIST  \\
43  &   0.28 $\pm$  0.09  &   0.89 $\pm$  0.10  &    0  &       37  &   \nodata &   \nodata &      K3.5  &     \nodata  &     \nodata  &        M4  &     \nodata  &     \nodata  & \nodata &             &                       UIST  \\
44  &   0.32 $\pm$  0.14  &   0.66 $\pm$  0.12  &    0  &       28  &   \nodata &   \nodata &      K1.5  &     \nodata  &     \nodata  &      K5.5  &     \nodata  &     \nodata  & \nodata &             &                       UIST  \\
45  &   0.34 $\pm$  0.09  &  $-$0.32 $\pm$  0.13  &    0  &       45  &   \nodata &   \nodata &        M0  &     \nodata  &     \nodata  &       $>$M7  &     \nodata  &     \nodata  & \nodata &             &                       UIST  \\
46  &   0.52 $\pm$  0.09  &   0.90 $\pm$  0.12  &    0  &       33  &   \nodata &   \nodata &        K3  &     \nodata  &     \nodata  &      M2.5  &     \nodata  &     \nodata  & \nodata &             &                       UIST  \\
47  &   0.51 $\pm$  0.09  &   0.78 $\pm$  0.12  &    0  &       24  &   \nodata &   \nodata &      K0.5  &     \nodata  &     \nodata  &        K4  &     \nodata  &     \nodata  & \nodata &             &                       UIST  \\
48  &   0.46 $\pm$  0.09  &   0.85 $\pm$  0.13  &    0  &       29  &   \nodata &   \nodata &      K1.5  &     \nodata  &     \nodata  &        M0  &     \nodata  &     \nodata  & \nodata &             &                       UIST  \\
\hline
\end{tabular}

\begin{list}{}{} 
\item[{\bf Notes:}] 
($^a$) EW(CO)s from low-resolution SofI spectra were multiplied by a factor 1.4.~
($^b$) sp\_RSG: spectral type by assuming a supergiant class.~
($^c$) sp\_giant: spectral type by assuming a giant class.~
($^d$) SofIl= low-resolution mode of SofI;   SofIm= medium-resolution mode of SofI.~

\end{list}
}
\end{sidewaystable*}

\subsection{Spectroscopic classification}
\label{subsec-spectralclass}
\subsubsection{Late-type stars}
We defined  as  late-type stars  those stars with effective temperatures lower than 4500 K
(red giant, asymptotic giant branch stars (AGBs), and  RSGs). 
We define  a star as a RSG when it has a \Teff\ lower than 4500 K and a  luminosity of
$L/L_\odot>10^4$, which corresponds to an initial mass of $\ga 9$ \Msun\ \citep{ekstrom12}.
Late-type stars can easily be identified with $K$ band 
spectroscopy, because of their strong  CO band-head at 2.29 \um\ 
\citep[e.g.,][]{kleinmann86, wallace96,rayner09, ivanov04}.   
Near-infrared spectra of  late-type stars (mostly Mira AGBs) may  show continuum absorption  
by water. 
Water absorption  affects both ends of the $H$ band spectrum, 
from $\sim 1.4$ \um\ to $\sim 1.55$ \um,
and from $\sim 1.75$ \um\ to $\sim 1.80$ \um, and the blue side of the $K$ band
from 2.0 \um\  to 2.1 \um\ \citep[e.g.,][]{comeron04, blum03, frogel87, alvarez00,rayner09}.
Water absorption   in $H$ band is most easily detected, as   a  curved (pseudo)-continuum.
Highly variable water absorption is found   in  Mira AGBs because of their  pulsation \citep{matsuura02}.
In Fig. \ref{sofi.spectra}, the stellar continua of stars 1 and 24 clearly show  absorption by water.

For red giants and RSGs, the equivalent width of the CO band-head, EW(CO), can be used as
a temperature indicator because it  increases linearly with decreasing $T_{eff}$.
Red giants and RSGs follow two different relations  \citep{blum03, frogel87, figer06};
for a given temperature, RSGs  have stronger CO bands than red giant stars.
EW(CO)s  of Miras are variables, do not  correlate with  $T_{eff}$ values,
and   their EW(CO)s  may be  as large as  those of late RSGs \citep{blum03}. 
Water absorption   decreases with increasing stellar 
luminosity \citep[][]{blum03, frogel87}. All these considerations mean that  
combined information on water absorption 
and EW(CO) is useful for estimating  luminosity classes.

%CO equivalent widths
We measured the  EW(CO)s  using the feature and continuum regions that are specified
in Table \ref{lines}; we obtained spectral types for the targets  by comparing their EW(CO)s
 with those of reference spectra from the atlas of \citet{kleinmann86};
 the reference spectra  were smoothed   to match
 the  spectral resolution of the targets \citep[e.g.,][]{figer06}. 
Twelve targets were observed with both UIST and SOFI detectors;
by comparing their resulting EW(CO)s, the typical accuracy of spectral types is found within two subclasses;
star 24   is a Mira-like star that   went from $>M7$ (UIST run) 
to M0 type (SOFI run).
For spectra taken  with the medium-resolution mode of SofI, 
we  selected a narrower bandpass for the CO feature 
(see Table \ref{lines}).
A scaling factor of 1.4 was measured between the EW(CO)s from the medium-resolution SofI and
those  from UIST.
For  star 27, which was observed with both  the low and medium modes of SofI,  
we obtained a RSG-type of K4 and K2, respectively.  
The degeneracy between RSGs and red giants disappears for EW(CO) values larger than $\sim 43\pm 4$ \AA. Miras
stars, with their   erratic behaviors, may also have   EW(CO)s  larger than 43 \AA.
Stars  1, 2, 15, 24, and 28 in Table 
\ref{table.spec} have  EW(CO) values larger than 48; these stars are candidate   AGBs or RSGs.

\begin{figure}[!]
\begin{centering}
\resizebox{0.99\hsize}{!}{\includegraphics[angle=0]{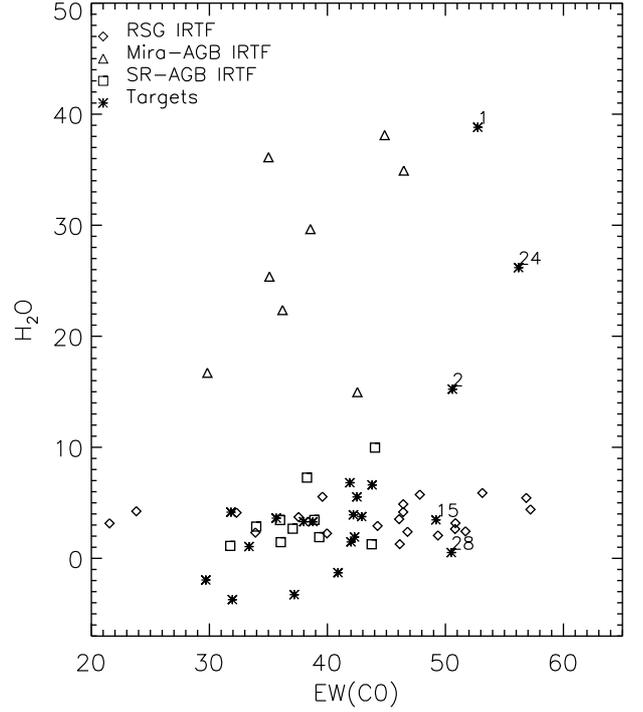}}
\end{centering}
\caption{\label{ewco} ${\mathrm H_{ 2} {\mathrm O}}$ indexes versus the EW(CO)s of targeted late-type stars and
 comparison samples of AGBs (Mira and semi-regular  variables) and RSGs from the IRTF library \citep{rayner09}. }
\end{figure}

\begin{table}
\caption{\label{lines} Spectroscopic indexes. Spectral regions used as features and continuum are  specified. }
%\begin{tabular}{lllll}
\begin{tabular}{@{\extracolsep{-.12in}}lllll}
\hline
\hline
Type   &       Band      & Continuum   & References    & Comment   \\
       &      [\um]      &  [\um]      &               &      \\
\hline\\
${\mathrm H_2\mathrm O}$ &  2.0525-2.0825 &  1.68-1.72   & 1 &\\ 
                         &               &   2.20-2.29   & 1 &  \\
CO     &       2.285-2.315&  2.28-2.29   &             &  UIST, SofI Low\\
CO     &       2.285-2.307&  2.28-2.29   &             & SofI medium\\
\hline\\
\end{tabular}
\begin{list}{}{}
\item[]{\bf References:} (1) \citet{blum03}.
\end{list}
\end{table}

%H2O index
Water indexes were obtained from the de-reddened low-resolution SofI spectra, which cover 
$H$ and $K$ bands. We used a water index with continuum and absorption  
regions defined  as in \citet{blum03}.  
We measured the water index  $(100 * (1-<F_{\mathrm H_2O}/F_{continuum}>))$  with a quadratic  fit
to two continuum regions (see Table \ref{lines}); $<F_{\mathrm H_2O}/F_{continuum}>$
is the average ratio of observed flux densities in the water region and 
estimated flux densities (with a fit).
For comparison, we estimated the EW(CO)s and water indexes
of a sample of known AGBs and RSGs  with spectra available from the IRTF  library  \citep{rayner09}.
In Fig.\ \ref{ewco}, we plot  water indexes versus  EW(CO)s of the
targets, as well as of reference spectra.
%Classification
A combination of water indexes and EW(CO)s  allows us to distinguish between RSGs and Mira AGBs; 
we find that known Mira AGBs have water indexes
greater than 15 \AA, as in  \citet{blum03}. This  confirms that  targets 1, 2, and 24 
are  Mira-like stars. Water indexes for  known  RSGs and semi-regular AGBs 
\citep[SR, e.g., ][]{alard01} are 
typically negligible;  SR AGBs  have EW(CO)s as large as $\sim44$ \AA; 
RSGs up to $\sim 58$ \AA\  (see Fig.\  \ref{ewco} and Table \ref{table.spec}).
SR  AGBs contaminate the sample of 
spectroscopic candidate RSGs (EW(CO)$ \ga 43$ \AA);
luminosities are crucial for classifying RSGs.

\subsubsection{Early-type stars}

Low-resolution infrared spectra of O-, B-, A-, and F-type stars are  characterized
by  hydrogen (H)  lines  \citep[e.g.,][]{hanson96,hanson98,lancon92, rayner09}.
The low-resolution SofI spectra of stars  9, 11, 13, 19, 20, 21, and 26
show  \ion{H}{I} lines in absorption at 1.555 \um, 1.570 \um, 1.588 \um,   1.611 \um, 
1.641 \um, 1.681 \um, 1.737 \um, and 2.166 \um\ 
(see Fig.\ \ref{sofi.spectra}). The  strengths of the hydrogen lines
indicate types from  B0 to F0 for stars 11, 13, 20, and 26; 
the presence of weak CO bands imply types from F5 to G5 for stars
9, 19, and 21.

A spectrum of star  39 was obtained with the medium-resolution mode SofI in $K$ band.
A \brgamma\ in absorption and a  \ion{He}{I} at 2.1127 \um\ are detected. The presence of the \ion{He}{I} line
implies   a B0-8 I, or a O9.5-B3V.

The coordinates and spectral types of  early-type stars are listed in Tables \ref{table.phot}
and \ref{table.spec}.

\subsubsection{Photometric variability index}
\label{subsec-photometric}
Mira AGBs are characterized by  periodic  photometric variations 
that can reach  up to  $\sim1$ mag in \Ks\ \citep{messineo04}. 
In contrast, only a small fraction of RSGs (about 20\%) are 
known to vary, and typically RSGs have amplitudes of a few tenths of a
magnitude in $K$ band  \citep{yang11}.

Since the two DENIS and 2MASS $J$ filters are  similar, we  used both measurements for identifying 
candidate variable stars, i.e., stars with $J$(DENIS)$ > 7.5$ mag, and $|J$ (DENIS)$ - J $(2MASS)$| > 3\times0.15$ mag, 
or with $K_{\mathrm s} $ (DENIS)$ > 6.0$ mag, and $|K_{\mathrm s} $ (DENIS)$ - K_{\mathrm s} $ (2MASS)$| > 3\times0.15$ mag.  
For nonvariable stars,  $|K_{\mathrm s} $ (DENIS)$ - K_{\mathrm s} $ (2MASS)$|$
and $|J $ (DENIS)$ - J $(2MASS)$|$ are within 0.15 mag \citep{schultheis00,messineo04}.
We also flagged as variables  those targets with indication of variability 
from the shortest WISE band.
The late-type stars   14, 17, 24, and 34 are 
candidate variables, as well as the early-types 13 and 20, and the $F$-$G$ stars 19 and 21. 
Star 24 is a Mira-like AGB, as inferred from the water index; the other  variable
late-type stars  could be SR AGBs \citep{alard01}.
None of the RSGs found has the variability flag on.

\subsubsection{ Target intrinsic colors and interstellar extinction}
\label{int.col}

 For red giants and RSGs,  we assumed the  intrinsic  $V-K$, $J-K$, and  $H-K$ colors per spectral-type 
 provided by   \citet{lejeune01} and \citet{koornneef83}. 
For each spectral type from K0 to M5, the $V-K$ (or  $J-K$)  colors of red giants and RSGs agree 
within 0.28 mag (or 0.09 mag), respectively \citep{koornneef83};   this quantity is  comparable to  
the color change between two 
neighboring subspectral types of the same luminosity class.
We obtained intrinsic  $I-J$  colors by interpolating a colored-isochrone of 6 Gyr and  
solar metallicity  \citep{pietrinferni04}. The  obtained $I-J$ colors of late-type  giants  
agree with  those of \citet{lejeune01} for  RSGs  (4-30 Myr old)  within 0.05 mag.
The Koornneef photometric system agrees with the 2MASS system within 0.09 mag \citep{carpenter01}. 

For early-type stars, infrared-colors  were taken from \citet{bibby08}, \citet{humphreys84}, \citet{wegner94},
\citet{koornneef83}, and \citet{wainscoat92}.

For every target, we estimated \Aks\ with  the assumed intrinsic colors, 
and a power-law extinction curve with an index of  $-1.9$ \citep{messineo05}.
This law yields  an excellent agreement between   estimates of \Aks\ from  the observed
 ($I$-$J$),  ($J-$\Ks), and ($H-$\Ks) colors. For eight stars with DENIS and 2MASS measurements, we obtained
$ {\it A_{\it K_{\rm s}}} ({\rm from~} J-K_{\rm s}) - {\it A_{\it K_{\rm s}}} ({\rm from~} I-J) =-0.04 ~{\rm mag}$ with $\sigma$=0.14 mag; 
$ {\it A_{\it K_{\rm s}}} ({\rm from~} J-K_{\rm s}) - {\it A_{\it K_{\rm s}}} ({\rm from~} H-K_{\rm s})=-0.03 ~{\rm mag}$ with $\sigma$=0.09 mag. 

 Typically, we adopted  the \Aks\  values from the ($J-$\Ks) colors.
For every field, an average  extinction ($<$\Aks$>$) was measured, 
as  the average of  the \Aks\ values of the observed late-type stars (with the exclusion of AGB stars).
The field results   are listed in Table \ref{table.targets}.
Stars with  a  value of \Aks\ that differs from the average extinction by more 
than three  standard deviations were classified as foreground (background) sources, as indicated in Table 
\ref{luminosity}. 
We used the shape of the continuum (quantified by the water index) 
to distinguish between objects with circumstellar envelopes (e.g., Mira AGB) and 
obscured, but naked, luminous objects.

\subsection{Interstellar extinction as a distance indicator}
\label{secdist}

Distances were derived  by matching the target \Aks\ value 
with a known curve of \Aks\ values versus distances (along the target line of sight).
The targets are  distributed  in five fields along  the Galactic plane, 
at galactic longitudes of 1$\rlap{.}^{\circ}$5, 9$\rlap{.}^{\circ}$5, 16$\rlap{.}^{\circ}$7, 
49$\rlap{.}^{\circ}3$, and 59$\rlap{.}^{\circ}8$, respectively, 
and approximately zero latitude. 

\begin{figure*}[!]
\begin{centering}
\resizebox{0.499\hsize}{!}{\includegraphics[angle=0]{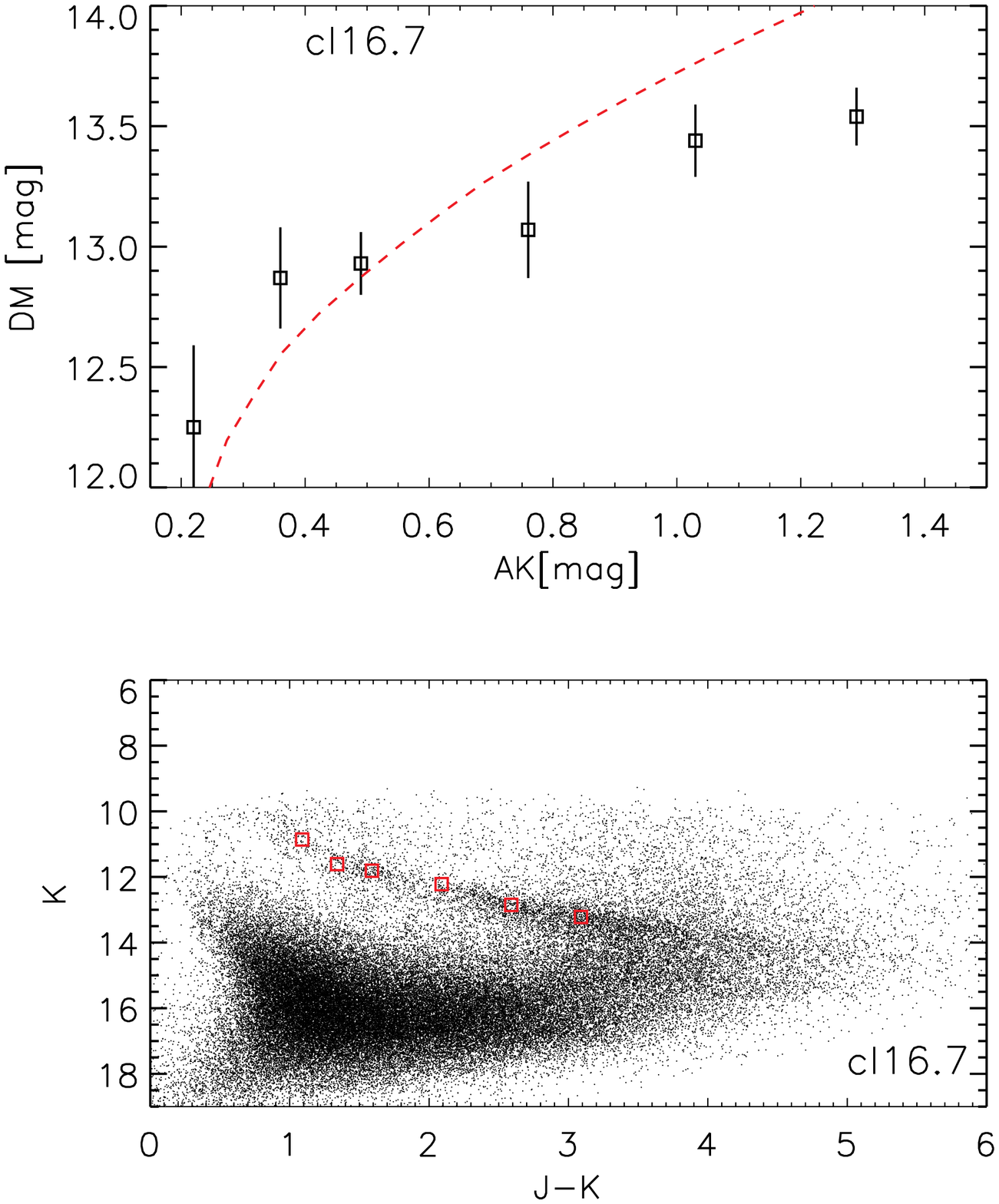}}
\resizebox{0.499\hsize}{!}{\includegraphics[angle=0]{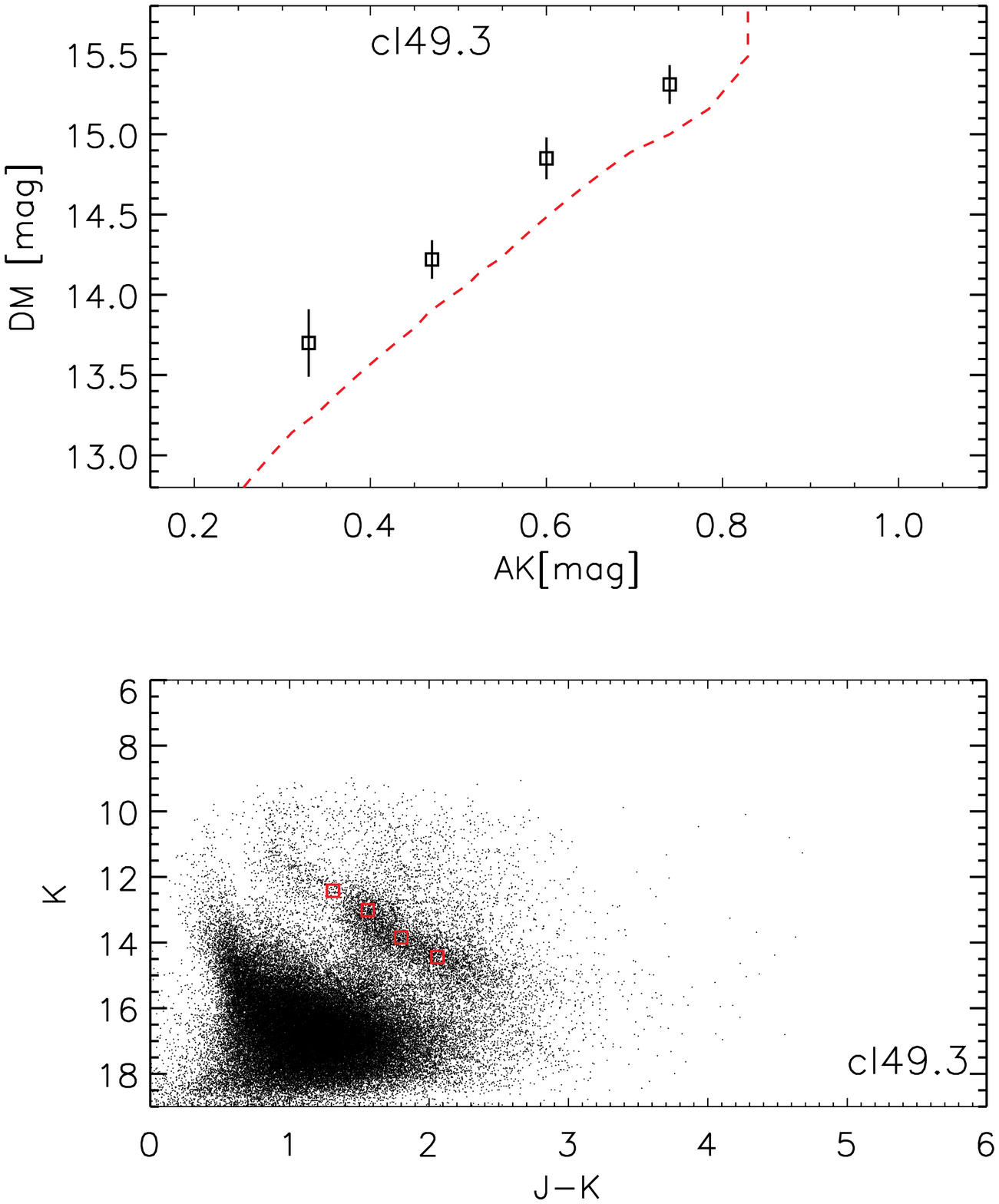}}
\end{centering}
\caption{\label{clump} {\it Top panels}: Interstellar extinction \Ak\  versus distance moduli of 
red clump stars in the fields of cl16.7 ({\it left}) and cl49.3 ({\it right}). As a comparison,
the model of \citet{drimmel03} is shown with a dashed line.
{\it Bottom panels}: UKIDSS $J-K$ versus $K$ diagram of data-points
within an area of $10$\arcmin$ \times 10$\arcmin\ from the
estimated centers of cl16.7 ({\it left}) and cl49.3 ({\it right}).
}
\end{figure*}

 A relation between  \Aks\ and distance along a given line of sight can be derived  
with primary distance calibrators.
Thanks to the availability of 
deep near-infrared surveys, such as VVV \citep{soto13} and UKIDSS \citep{lucas08},  
red clump stars (analogous to the  Galactic horizontal branch stars in
metal rich globular clusters) have been detected throughout the Galaxy, and may serve
for this purpose \citep[e.g.,][]{gonzalez11,drimmel03}. 
We created UKIDSS $K$ versus $J-K$ diagrams of datapoints within 
$10$\arcmin$\times10$\arcmin\ from the  cluster centers, and
we visually selected by eyes the locus of a well visible  red clump sequence.
Per  bin of $J-K$ color, we  analyzed the distribution of  $K$ magnitudes 
(around the clump region), and estimated  the peak of the red clump with a Gaussian fit. 
We assumed an intrinsic $(J-K)_o$ of 0.68 mag  \citep{babusiaux05,gonzalez11} and an absolute  
magnitude in $K$ band, \Mk, of $-1.61$ mag  \citep{alves00}. Quoted values  in  the literature 
range from \Mk=$-1.54,-1.55$ mag \citep{groenewegen08,pietrinferni04} 
to \Mk=$-1.72$  mag \citep{babusiaux05}.
For all fields but one,  distances were derived with clump stars in the \Aks\ region
of interest (average \Aks\ of the targeted stars); for field cl59.8
it was not possible because   there were not enough  clump stars.
For two fields, cl16.7 and cl49.3,   
it was possible to derived a
curve of variation of \Aks\ with distance \citep[e.g.,][]{drimmel03}, 
 as shown  in Fig.\ \ref{clump}. 

Alternatively, a curve of \Aks\ versus distance along a given line of sight was derived with a
model of  Galactic dust distribution from COBE/DIRBE 
\citep{drimmel03,drimmel01}.
A scaling factor of 0.74 was applied to transform the \Aks\ of \citet{drimmel03} 
and \citet{rieke85} into the \Aks\ derived with a  power law of index=$-1.9$ \citep{messineo05}.

Estimated  distances are listed in Table \ref{table.targets}.
In the following, we  assume the  distances derived with red clump stars.  
The Drimmel model tends to give shorter 
distances than those from the red clump  fitting (Fig.\ \ref{clump});
however, any detected RSGs would still be a
RSG at the smaller distance provided by the Drimmel model.

In field cl49.3 it was possible to derive a spectrophotometric distance for star 39,
an early B star with  \Ks=11.19 mag. 
Typically, early-B stars have an absolute $K$ of 
$-6.45\pm0.42$ mag,  $-3.52\pm1.00$ mag, or $-2.38\pm0.77$ mag,
when they are supergiants (Ia), giants, or dwarfs, respectively
\citep{humphreys84,wegner94,bibby08,koornneef83}.
A supergiant class and a distance of 26 kpc can be excluded. 
By assuming a dwarf or a giant class, we derived 
a distance modulus (DM) of  $13.02\pm0.77$ mag (4.0 kpc) and $14.16\pm1.00$ mag (6.8 kpc), respectively.

\subsection{Estimated luminosities}
\label{subsection-luminosities}
Bolometric magnitudes, \Mbol, and luminosities were calculated with  \Ks\ magnitudes,
 \Aks, estimated DMs, and bolometric corrections, \BCK\
(see  Table  \ref{luminosity}). Values of \BCK\   per spectral type
were taken from  \citet[][]{levesque05} and \citet{messineo11};
the solar \Mbol\ was assumed to be +4.74 \citep{bessell98}.

For late-type stars, the spectral-energy distribution (\Teff\ from 4500 to 2000  K) peaks in the 
near-infrared (0.6-1.5 \um), or longward of this range when  circumstellar dust is present 
\citep[e.g.,][]{ortiz02,figer06}. For late-types, we additionally integrated available 
flux densities over frequency $\nu$ using a linear interpolation. 
At short wavelengths, flux densities were extrapolated to zero  with a blackbody function,
at long wavelengths with a linear   interpolation through the last two points. 
Direct integration yields more robust luminosities for   dusty AGB stars, 
for which \BCK\ of naked stars are not applicable \citep{messineothesis,ortiz02}. 
The solar constant, $C_{\rm bol}$, was assumed to be $-18.90$.
Our integration  method provided  \Mbol\ values in agreement  within errors with those 
derived  with the  \BCK\ values.

Known RSGs have typical \Mbol\ from $-5.5$ mag to $ -9.0$ mag 
\citep{figer06,davies07,clark09}.  Galactic mass-losing AGB stars rarely
exceed \Mbol$\approx-7.1$ mag \citep{wood93};  
this theoretical limit may be exceeded with the onset of the hot bottom burning (HBB) process
that breaks down the classic  relation between  core mass and   luminosity 
\citep[see, e.g., ][]{groenewegen09, paczynski71}. 
However, the efficiency of the HBB mechanism anti-correlates with metallicity,
and  it is usually  associated with a stellar regime of high mass-loss and superwind 
\citep[e.g., ][]{wood83, marigo01,wood93};  typically, AGBs have   
strong water absorption
in the bright end  \citep[\Mbol\ from $-5.5$ to  $-7$ mag,][]{habing03,alard01}. 
In the Milky Way, Mira AGBs have  mostly late-types  
\citep[M7-M9, as reported, for example, by ][]{rayner09};  the distribution of spectral types  of 
known RSGs   peaks at M2-M4  \citep{davies07}.

For selecting clusters of RSGs, we 
calculated the photometric properties of the targets at the distance provided by 
their average \Aks\ per field (as explained  in Sect.\ \ref{secdist});
after having selected  candidate RSGs, 
the procedure was reiterated with the \Aks\ of the candidate RSGs. 
The results are summarized in Table \ref{luminosity}.
In field cl16.7 (DM=12.9 mag), stars 22 (K4.5-5I) and 23 (K5.5I) are two  RSGs 
with \Mbol=$-6.18$  mag and $-5.71$ mag. In field cl49.3 (DM=14.2 mag), star 27 (K4I) with 
a \Mbol=$-7.82$ mag is a RSG star, along with the nearby 
stars 28 (M1-1.5I), 29 (K5.5-M0I),  and 30 (K5-K5.5I), which have 
\Mbol=$-6.57$ mag, $-6.07$ mag, and $-5.76$ mag, respectively.
In none of these stars did we detect  signs of variability or  water absorption.
There is only one Mira-like star exceeding \Mbol=$-5.5$ mag in field cl1.5, which is  star 1.
For all other late-type stars, we estimated \Mbol\ typical of giant stars. 
There are a number of giants (after excluding the detected Mira-like AGBs) with 
estimated types later than M7III (3,  5, 12, 14, 15, 42, and 45); 
their \Mbol\  are  below the values expected for faint  RSGs.

\section{Lines of sight  towards cl16.7 and cl49.3}
\label{offthere}

\begin{figure}[!]
\begin{centering}
\resizebox{\hsize}{!}{\includegraphics[angle=0]{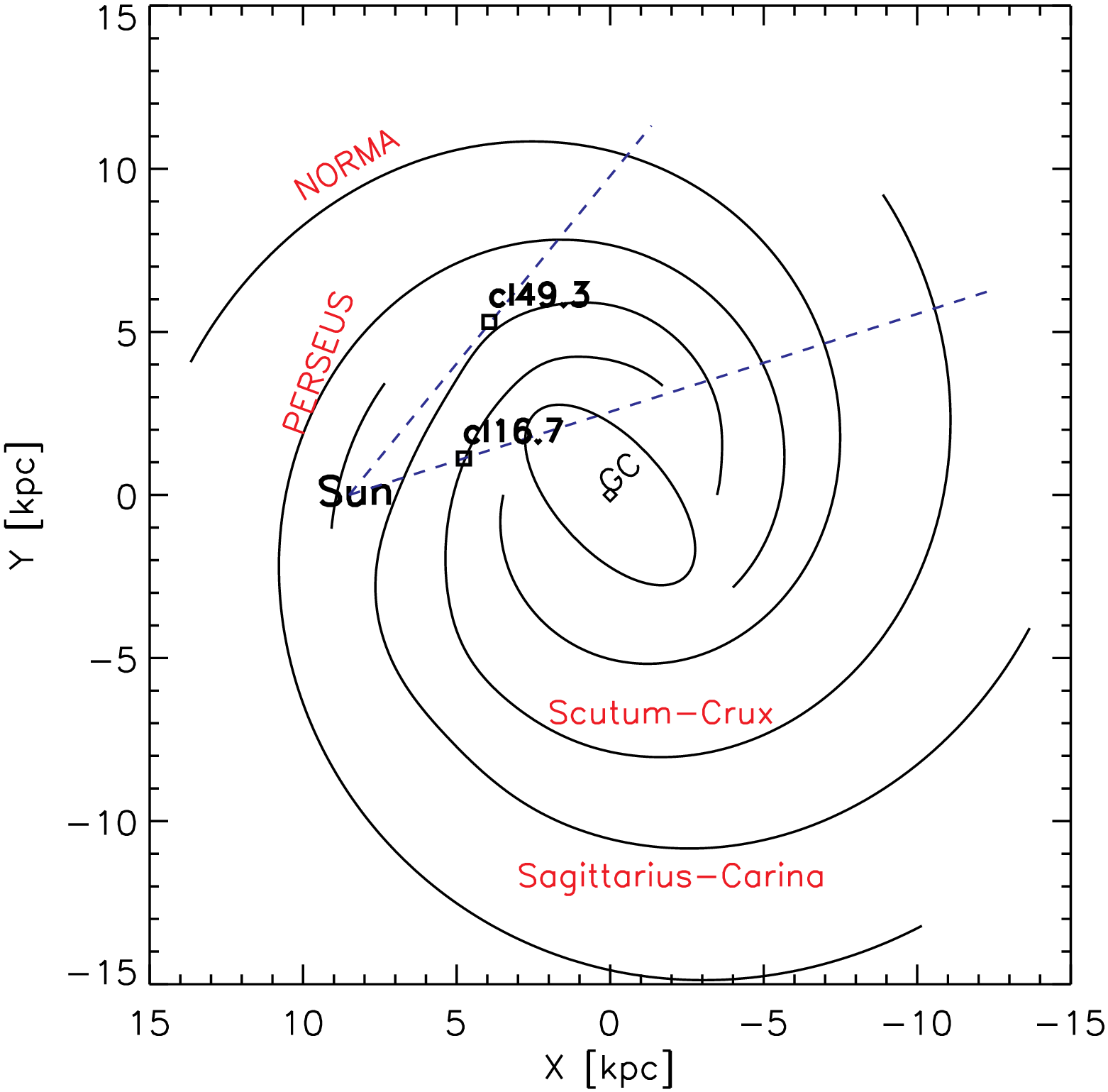}}
\end{centering}
\caption{\label{tondi}  
XY view of the Milky Way. The Galactic center is at (0,0) and the Sun is at  (8,0). 
The dashed lines indicate the line of sights to cl16.7 and cl49.3, which are marked
with squares. A sketch of the spiral structure is also drawn and the arms are labeled
\citep{lazio02}. The X-axis is oriented along the Sun-Galactic center line of sight. }
\end{figure}

 In the CMD of field cl16.7, at 16$\rlap{.}^{\circ}$75  of longitude, 
and  $-0\rlap{.}^{\circ}$63 of latitude, the red clump trace is clearly visible  from  $J-K\approx0.8$ mag
and $K\approx10$ mag to $J-K\approx3.2$ mag  and $K\approx13.5$ mag (see Fig.\ \ref{clump}). 
Clump stars appear
evenly distributed along the line of sight, and their number  increases with 
distance. For the RSGs in cl16.7, we derived an average \Aks\ of 0.49 mag;
from this \Aks, we obtained a DM of 12.93 mag, which   places them on the Scutum-Crux 
arm  \citep[see Fig.\  \ref{tondi},  as well as the distribution of Galactic massive clusters in ][]{messineo09,messineo14}.

Field  cl49.3 is located at a longitude of 49$\rlap{.}^{\circ}$34 and a latitude of +0$\rlap{.}^{\circ}$72.
The DM inferred from the RSGs places the cluster cl49.3 
onto the Sagittarius-Carina arm (see Fig.\  \ref{tondi}).
In the CMD of field cl49.3, red clump stars are traceable from $J-K\approx1$ mag and  $K\approx11.5$ mag
to $J-K\approx2.1$ mag and  $K\approx14.5$ mag (see Fig.\ \ref{clump}). There is a sudden increase in the number
of red clump stars at $J-K \approx 1.4 $ mag, and a heliocentric distance larger than  $\sim 5$  kpc.
This overdensity of red clump stars occurs at the shortest Galacto-centric distances along the line of sight, 
where the stellar density is higher; they are also encountered when  tangentially approaching the Sagittarius-Carina arm 
(heliocentric distance 5-9 kpc, see Fig.\  \ref{tondi}).
Similar structures appear in the CMDs of neighboring bins of longitudes.
Some models predict the Galactic outer Linbland resonance to occur inside the solar circle, 
at a Galacto-centric distance of about 5 kpc \citep[e.g.,][]{habing06};  
such resonances may generate/enforce extended structures  \citep[e.g., rings and pseudo-rings,][]{buta95}. 
The Galactic long bar only  extends  to about $\la 30$\degr\   longitude \citep{cabrera08}.

\begin{table*}
\caption{ \label{luminosity}
Photometric properties of the targeted stars. }
\begin{tabular}{@{\extracolsep{-.12in}}rrlrlrrrrrrrrrrrrr}
\hline
\hline
{\rm ID}  & {\rm rad }  & {\rm Fields}   & {\it K$_{\rm S}$ } &  {\rm Sp$^*$} & {\it (J$-$K$_{\rm S})_o$ } &
{ ${\it A}_{\it K_{\rm s}}$(\it JK${_{\rm s}}$)} & { ${\it A}_{\it K_{\rm s}}$(\it HK${_{\rm s}}$)}&{ ${\it A}_{\it  K_{\rm s}}$(\it IJ)} &   ${\it M}_{\it K_{\rm}}$ &DM&$BC_{\it K}$ & ${\it M_{\rm bol1}}$ & ${\it M_{\rm bol2}}$& Notes$^{**}$  \\
          & [\arcmin]  &      & [mag] &   & [mag] &[mag] & [mag] & [mag] &   [mag] &[mag]&[mag] & [mag] & [mag]& \\
 \hline

   1 &  3.30 &      cl1.5 &  4.86 &       M8IIIM &  1.42 & 0.64 & 0.89 & \nodata  & $$-$$10.32 $\pm$  0.17 & 14.54 $\pm$  0.16 &  3.34 $\pm$  0.11 & $$-$$6.98 $\pm$  0.21 & $$-$$7.55 $\pm$  0.33 &       \\
   2 &  0.09 &      cl1.5 &  7.35 &       M8IIIM &  1.42 & 0.49 & 0.58 & \nodata  &  $$-$$7.68 $\pm$  0.17 & 14.54 $\pm$  0.16 &  3.34 $\pm$  0.11 & $$-$$4.33 $\pm$  0.21 & $$-$$4.59 $\pm$  0.32 &       \\
   3 &  0.37 &      cl1.5 &  8.09 &        M8III &  1.42 & 0.57 & 0.58 & \nodata  &  $$-$$7.02 $\pm$  0.18 & 14.54 $\pm$  0.16 &  3.34 $\pm$  0.11 & $$-$$3.67 $\pm$  0.21 & $$-$$3.92 $\pm$  0.33 &       \\
   4 &  0.24 &      cl1.5 &  8.54 &      M6.5III &  1.33 & 0.38 & 0.44 & \nodata  &  $$-$$6.38 $\pm$  0.18 & 14.54 $\pm$  0.16 &  3.15 $\pm$  0.11 & $$-$$3.23 $\pm$  0.21 & $$-$$3.39 $\pm$  0.36 &       \\
   5 &  1.81 &      cl1.5 &  8.81 &        M8III &  1.42 & 0.32 & 0.28 & \nodata  &  $$-$$6.05 $\pm$  0.19 & 14.54 $\pm$  0.16 &  3.34 $\pm$  0.11 & $$-$$2.71 $\pm$  0.22 & $$-$$3.00 $\pm$  0.40 &       \\
   6 &  0.10 &      cl1.5 &  9.01 &      M1.5III &  1.02 & 0.55 & 0.51 & \nodata  &  $$-$$6.08 $\pm$  0.17 & 14.54 $\pm$  0.16 &  2.76 $\pm$  0.10 & $$-$$3.32 $\pm$  0.20 & $$-$$3.42 $\pm$  0.32 &       \\
   7 &  2.92 &      cl1.5 &  9.07 &      M3.5III &  1.13 & 0.40 & 0.51 & \nodata  &  $$-$$5.87 $\pm$  0.18 & 14.54 $\pm$  0.16 &  2.88 $\pm$  0.11 & $$-$$2.99 $\pm$  0.21 & $$-$$3.08 $\pm$  0.36 &       \\
   8 &  4.15 &      cl1.5 & 10.06 &      M4.5III &  1.20 & \nodata  & 0.53 & \nodata  &      \nodata  & 14.54 $\pm$  0.16 &  2.89 $\pm$  0.11 &     \nodata  & $$-$$2.43 $\pm$  0.21 &       \\
   9 &  2.04 &      cl1.5 & 10.08 &         F5G5 &  0.34 & 0.31 & 0.22 & \nodata  &  $$-$$4.77 $\pm$  0.16 & 14.54 $\pm$  0.16 &     \nodata  &     \nodata  &     \nodata  &       \\
  10 &  3.68 &      cl1.5 & 10.23 &      M2.5III &  1.07 & 0.46 & 0.45 & \nodata  &  $$-$$4.76 $\pm$  0.17 & 14.54 $\pm$  0.16 &  2.83 $\pm$  0.09 & $$-$$1.93 $\pm$  0.20 & $$-$$2.06 $\pm$  0.35 &       \\
  11 &  0.43 &      cl1.5 & 12.24 &         B0F0 & $$$-$$$0.06 & 0.16 & 0.09 & \nodata  &   0.89 $\pm$   0.80 & 11.19 $\pm$  0.80 &     \nodata  &     \nodata  &     \nodata  &    FG \\
  12 &  0.24 &      cl9.5 &  6.39 &        M7III &  1.36 & 0.63 & 0.60 & 0.55 &  $$-$$8.29 $\pm$  0.14 & 14.05 $\pm$  0.12 &  3.21 $\pm$  0.11 & $$-$$5.08 $\pm$  0.18 & $$-$$5.25 $\pm$  0.32 &       \\
  13 &  0.21 &      cl9.5 &  7.83 &   B0F0/A0III &  0.09 & 0.04 & 0.17 & \nodata  &   $-$1.10 $\pm$   0.80 &  8.90 $\pm$  0.80 &     \nodata  &     \nodata  &     \nodata  &    FG \\
  14 &  0.27 &      cl9.5 &  8.02 &        M7III &  1.36 & 0.73 & 0.74 & 0.83 &  $$-$$6.76 $\pm$  0.14 & 14.05 $\pm$  0.12 &  3.21 $\pm$  0.11 & $$-$$3.55 $\pm$  0.18 & $$-$$3.69 $\pm$  0.31 &       \\
  15 &  0.16 &      cl9.5 &  8.13 &        M8III &  1.42 & 0.73 & 0.72 & \nodata  &  $$-$$6.65 $\pm$  0.14 & 14.05 $\pm$  0.12 &  3.34 $\pm$  0.11 & $$-$$3.30 $\pm$  0.18 & $$-$$3.55 $\pm$  0.29 &       \\
  16 &  0.29 &      cl9.5 &  8.84 &      M0.5III &  0.99 & 0.87 & 0.87 & 0.84 &  $$-$$6.08 $\pm$  0.13 & 14.05 $\pm$  0.12 &  2.71 $\pm$  0.10 & $$-$$3.36 $\pm$  0.17 & $$-$$3.45 $\pm$  0.26 &       \\
  17 &  1.05 &      cl9.5 &  9.49 &        M4III &  1.16 & 0.70 & 0.88 & 0.64 &  $$-$$5.27 $\pm$  0.14 & 14.05 $\pm$  0.12 &  2.89 $\pm$  0.11 & $$-$$2.38 $\pm$  0.18 &  \nodata $\pm$  \nodata  &       \\
  18 &  1.30 &      cl9.5 &  9.60 &      M1.5III &  1.02 & 0.59 & 0.54 & 0.67 &  $$-$$5.04 $\pm$  0.13 & 14.05 $\pm$  0.12 &  2.76 $\pm$  0.10 & $$-$$2.28 $\pm$  0.17 & $$-$$2.37 $\pm$  0.30 &       \\
  19 &  1.81 &      cl9.5 & 11.76 &         F5G5 &  0.34 & 0.01 & 0.07 & \nodata  &   5.82 $\pm$   0.80 &  5.93 $\pm$   0.80 &     \nodata  &     \nodata  &     \nodata  &    FG \\
  20 &  2.37 &      cl9.5 & 11.77 &         B0F0 & $$$-$$$0.06 & 0.36 & 0.40 & \nodata  &  $$-$$2.65 $\pm$     0.13 & 14.05 $\pm$  0.12 &     \nodata  &     \nodata  &     \nodata  &       \\
  21 &  0.92 &      cl9.5 & 12.49 &         F5G5 &  0.34 & 0.75 & 0.58 & \nodata  &  $$-$$2.30 $\pm$  0.13 & 14.05 $\pm$  0.12 &     \nodata  &     \nodata  &     \nodata  &       \\
  22 &  0.52 &     cl16.7 &  4.55 &          K5I &  0.96 & 0.46 & 0.59 & 0.54 &  $$-$$8.84 $\pm$  0.15 & 12.93 $\pm$  0.13 &  2.66 $\pm$  0.10 & $$-$$6.18 $\pm$  0.18 & $$-$$6.24 $\pm$  0.36 &       \\
  23 &  0.38 &     cl16.7 &  5.06 &        K5.5I &  0.98 & 0.52 & 0.62 & 0.87 &  $$-$$8.39 $\pm$  0.15 & 12.93 $\pm$  0.13 &  2.68 $\pm$  0.10 & $$-$$5.71 $\pm$  0.18 & $$-$$5.79 $\pm$  0.35 &       \\
  24 &  0.71 &     cl16.7 &  5.92 &       M8IIIM &  1.42 & 1.29 & 1.49 & 0.24 &  $$-$$8.31 $\pm$  0.15 & 12.93 $\pm$  0.13 &  3.34 $\pm$  0.11 & $$-$$4.97 $\pm$  0.18 & $$-$$5.37 $\pm$  0.30 &       \\
  25 &  0.61 &     cl16.7 &  6.64 &        M6III &  1.30 & 0.63 & 0.65 & 0.52 &  $$-$$6.91 $\pm$  0.15 & 12.93 $\pm$  0.13 &  3.08 $\pm$  0.11 & $$-$$3.83 $\pm$  0.18 & $$-$$3.98 $\pm$  0.31 &       \\
  26 &  1.10 &     cl16.7 & 12.10 &         B0F0 & $$$-$$$0.06 & 0.58 & 0.66 & \nodata  &  $$-$$1.41 $\pm$   0.15 & 12.93 $\pm$  0.13 &     \nodata  &     \nodata  &     \nodata  &       \\
  27 &  0.59 &     cl49.3 &  4.12 &          K4I &  0.79 & 0.31 & 0.34 & \nodata  & $$-$$10.41 $\pm$  0.46 & 14.22 $\pm$  0.12 &  2.59 $\pm$  0.10 & $$-$$7.82 $\pm$  0.47 & $$-$$7.99 $\pm$  0.74 &       \\
  28 &  0.92 &     cl49.3 &  5.41 &        M1.5I &  1.03 & 0.51 & 0.51 & \nodata  &  $$-$$9.32 $\pm$  0.14 & 14.22 $\pm$  0.12 &  2.76 $\pm$  0.10 & $$-$$6.57 $\pm$  0.17 & $$-$$6.66 $\pm$  0.31 &       \\
  29 &  1.45 &     cl49.3 &  5.97 &        K5.5I &  0.98 & 0.50 & 0.54 & \nodata  &  $$-$$8.75 $\pm$  0.14 & 14.22 $\pm$  0.12 &  2.68 $\pm$  0.10 & $$-$$6.07 $\pm$  0.17 & $$-$$6.15 $\pm$  0.34 &       \\
  30 &  0.72 &     cl49.3 &  6.38 &          K5I &  0.96 & 0.57 & 0.53 & \nodata  &  $$-$$8.42 $\pm$  0.15 & 14.22 $\pm$  0.12 &  2.66 $\pm$  0.10 & $$-$$5.76 $\pm$  0.18 & $$-$$5.84 $\pm$  0.35 &       \\
  31 &  3.59 &     cl49.3 &  7.16 &        K3III &  0.80 & 0.21 & 0.25 & \nodata  &  $$-$$7.27 $\pm$  0.14 & 14.22 $\pm$  0.12 &  2.53 $\pm$  0.10 & $$-$$4.74 $\pm$  0.17 & $$-$$4.86 $\pm$  0.35 &       \\
  32 &  3.86 &     cl49.3 &  7.40 &      K5.5III &  0.97 & 0.97 & 1.31 & \nodata  &  $$-$$7.79 $\pm$  \nodata  & 14.22 $\pm$  0.12 &  2.68 $\pm$  0.10 & $$-$$5.11 $\pm$  \nodata  & $$-$$5.25 $\pm$  0.28 &       \\
  33 &  5.10 &     cl49.3 &  7.92 &        K5III &  0.96 & 1.18 & 1.47 & \nodata  &  $$-$$7.48 $\pm$  \nodata  & 14.22 $\pm$  0.12 &  2.66 $\pm$  0.10 & $$-$$4.82 $\pm$  \nodata  & $$-$$4.95 $\pm$  0.31 &       \\
  34 &  3.05 &     cl49.3 &  8.06 &        K2III &  0.68 & 0.21 & 0.22 & \nodata  &  $$-$$6.38 $\pm$  0.14 & 14.22 $\pm$  0.12 &  2.46 $\pm$  0.03 & $$-$$3.91 $\pm$  0.14 & $$-$$4.08 $\pm$  0.33 &       \\
  35 &  5.02 &     cl49.3 &  8.09 &        K0III &  0.64 & 0.07 & 0.10 & \nodata  &  $$-$$6.20 $\pm$  0.14 & 14.22 $\pm$  0.12 &  2.54 $\pm$  0.03 & $$-$$3.66 $\pm$  0.14 & $$-$$3.90 $\pm$  0.32 &       \\
  36 &  1.43 &     cl49.3 &  8.39 &      K3.5III &  0.84 & 0.52 & 0.46 & \nodata  &  $$-$$6.35 $\pm$  0.14 & 14.22 $\pm$  0.12 &  2.56 $\pm$  0.10 & $$-$$3.79 $\pm$  0.17 & $$-$$3.90 $\pm$  0.31 &       \\
  37 &  1.12 &     cl49.3 & 10.07 &      K3.5III &  0.84 & 0.59 & 0.54 & \nodata  &  $$-$$4.74 $\pm$  0.14 & 14.22 $\pm$  0.12 &  2.56 $\pm$  0.10 & $$-$$2.18 $\pm$  0.17 & $$-$$2.29 $\pm$  0.32 &       \\
  38 &  1.31 &     cl49.3 & 10.24 &        M0III &  0.97 & 0.83 & 0.84 & \nodata  &  $$-$$4.81 $\pm$  0.13 & 14.22 $\pm$  0.12 &  2.70 $\pm$  0.10 & $$-$$2.11 $\pm$  0.16 & $$-$$2.19 $\pm$  0.27 &       \\
  39 &  0.86 &     cl49.3 & 11.19 &      O9.5-B3 & $-$0.13 & 0.56 & 0.51 & \nodata  &   $-$3.58 $\pm$  0.12 &  14.22 $\pm$  0.12 & $-$2.53 $\pm$  0.87 & $-$6.11 $\pm$  0.88 &     \nodata  &       \\
  40 &  5.80 &     cl49.3 & 11.29 &        K3III &  0.80 & 0.78 & 0.81 & \nodata  &  $$-$$3.71 $\pm$  0.15 & 14.22 $\pm$  0.12 &  2.53 $\pm$  0.10 & $$-$$1.19 $\pm$  0.18 &  \nodata $\pm$  \nodata  &       \\
  41 &  1.36 &     cl49.3 & 11.48 &        K0III &  0.64 & 0.24 & 0.23 & \nodata  &  $$-$$2.98 $\pm$  0.64 & 14.22 $\pm$  0.12 &  2.54 $\pm$  0.03 & $$-$$0.44 $\pm$  0.64 &  \nodata $\pm$  \nodata  &       \\
  42 &  0.34 &     cl59.8 &  6.49 &        M8III &  1.42 & 0.56 & 0.61 & \nodata  &  $$-$$6.92 $\pm$  1.39 & 12.85 $\pm$  1.39 &  3.34 $\pm$  0.11 & $$-$$3.58 $\pm$  1.40 & $$-$$3.82 $\pm$  1.56 &       \\
  43 &  0.26 &     cl59.8 &  6.73 &        M4III &  1.16 & 0.37 & 0.43 & \nodata  &  $$-$$6.49 $\pm$  1.39 & 12.85 $\pm$  1.39 &  2.89 $\pm$  0.11 & $$-$$3.60 $\pm$  1.40 & $$-$$3.69 $\pm$  1.56 &       \\
  44 &  0.28 &     cl59.8 &  6.97 &      K5.5III &  0.97 & 0.06 & 0.09 & \nodata  &  $$-$$5.93 $\pm$  1.39 & 12.85 $\pm$  1.39 &  2.68 $\pm$  0.10 & $$-$$3.25 $\pm$  1.40 & $$-$$3.35 $\pm$  1.54 &       \\
  45 &  0.72 &     cl59.8 &  8.22 &        M8III &  1.42 & 0.79 & 0.72 & \nodata  &  $$-$$5.42 $\pm$  \nodata  & 12.85 $\pm$  1.39 &  3.34 $\pm$  0.11 & $$-$$2.07 $\pm$  \nodata  & $$-$$2.36 $\pm$  1.56 &       \\
  46 &  0.47 &     cl59.8 &  8.39 &      M2.5III &  1.07 & 0.42 & 0.36 & \nodata  &  $$-$$4.88 $\pm$  1.39 & 12.85 $\pm$  1.39 &  2.83 $\pm$  0.09 & $$-$$2.05 $\pm$  1.40 & $$-$$2.17 $\pm$  1.56 &       \\
  47 &  0.84 &     cl59.8 &  8.54 &        K4III &  0.88 & 0.11 & 0.04 & \nodata  &  $$-$$4.42 $\pm$  1.39 & 12.85 $\pm$  1.39 &  2.59 $\pm$  0.10 & $$-$$1.83 $\pm$  1.40 & $$-$$1.94 $\pm$  1.59 &       \\
  48 &  0.95 &     cl59.8 &  9.03 &        M0III &  0.97 & 0.36 & 0.32 & \nodata  &  $$-$$4.17 $\pm$  1.39 & 12.85 $\pm$  1.39 &  2.70 $\pm$  0.10 & $$-$$1.48 $\pm$  1.40 & $$-$$1.59 $\pm$  1.53 &       \\
\hline
\end{tabular}

\begin{list}{}{}
\item[{\bf Notes:}] Identification numbers are followed by  distances to the centers of the stellar overdensities,
 field names, \Ks,   spectral types, ($J-$\Ks)$_o$,
\Aks ($J-$\Ks), \Aks ($H-$\Ks), \Aks ($I-J$), absolute \Ks\ (\Mk), distance modulus (DM), bolometric corrections (\BCK), 
and bolometric magnitudes.
\Mbolone\ are obtained via \BCK, and \Mboltwo\ via direct integration.~
($^*$) The M suffix indicates a Mira-like star.~ 
($^{**}$) The FG  comment indicates that this star has an \Aks\ at least three times smaller than the average of the targets in the field.
 Its distance was estimated with Drimmel model (a significant number of clump stars were not available at this \Aks).
The quoted DM error (0.8 mag) was estimated from Fig.\ 6.~ 
\end{list}

\end{table*}

\section{Clusterings}
\label{secclusterings}

Because of their young age most RSGs are mostly found in stellar clusters. 
Massive young clusters rich in RSGs have been  identified via their near-infrared 
CMDs, e.g., the RSGC1 \citep{figer06}, RSGC2 \citep{davies07}, and RSGC3 \citep{clark09}.
In their ($J-$\Ks) versus \Ks\ diagrams there are gaps of several magnitudes in \Ks\ 
between RSGs and blue supergiants \citep{figer06}; 
the gap inversely correlates with the cluster age \citep{figer06,messineo09}.
For comparison, Mira AGBs are typically isolated on a scale of 1\arcmin\ and 
are among the brightest stars, but are typically the reddest \citep{messineo05,messineo04};
in the CMDs of stars surrounding Mira stars gaps may only be present stochastically.

\begin{figure*}[!]
\begin{centering}
\resizebox{0.5\hsize}{!}{\includegraphics[angle=0]{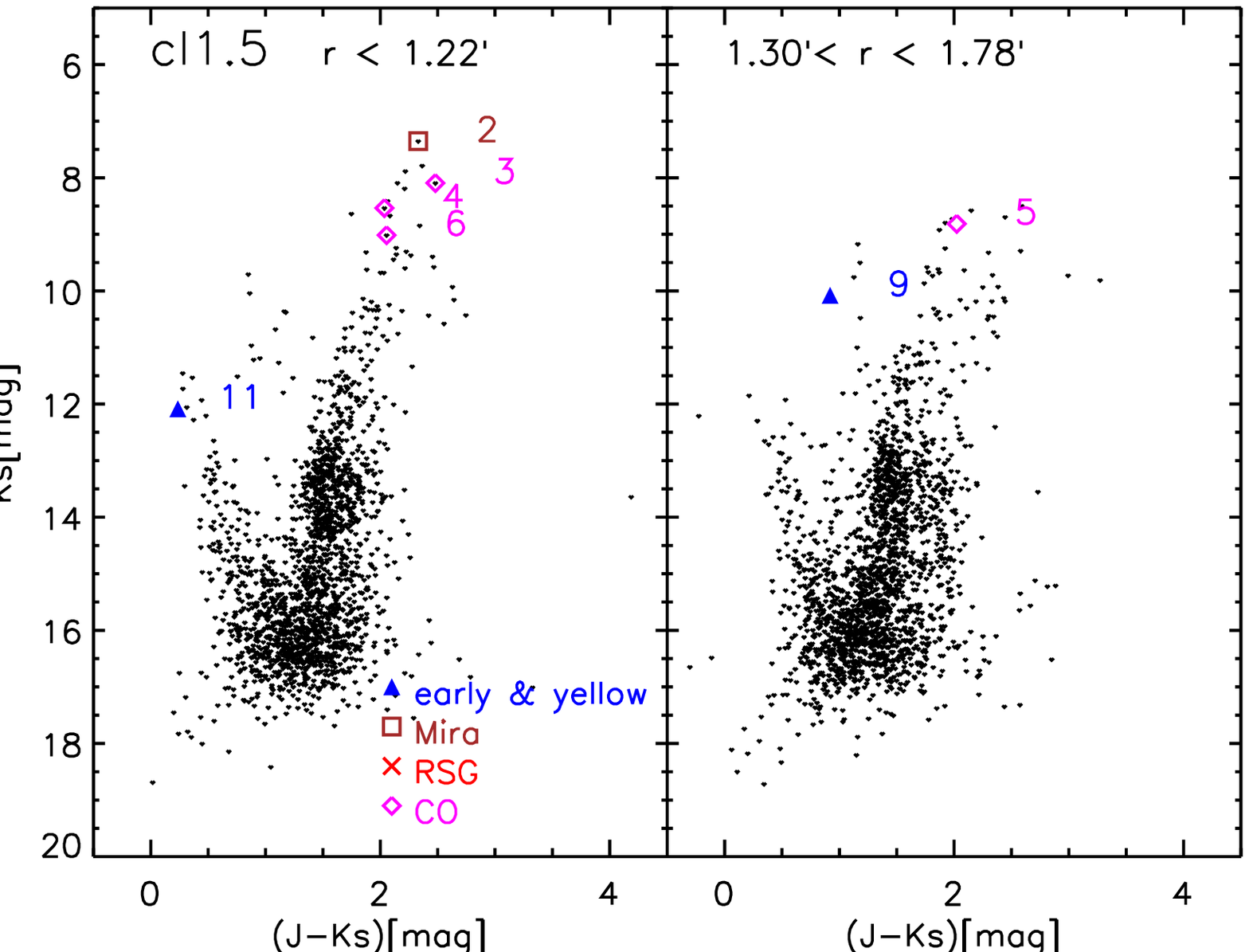}}
\resizebox{0.5\hsize}{!}{\includegraphics[angle=0]{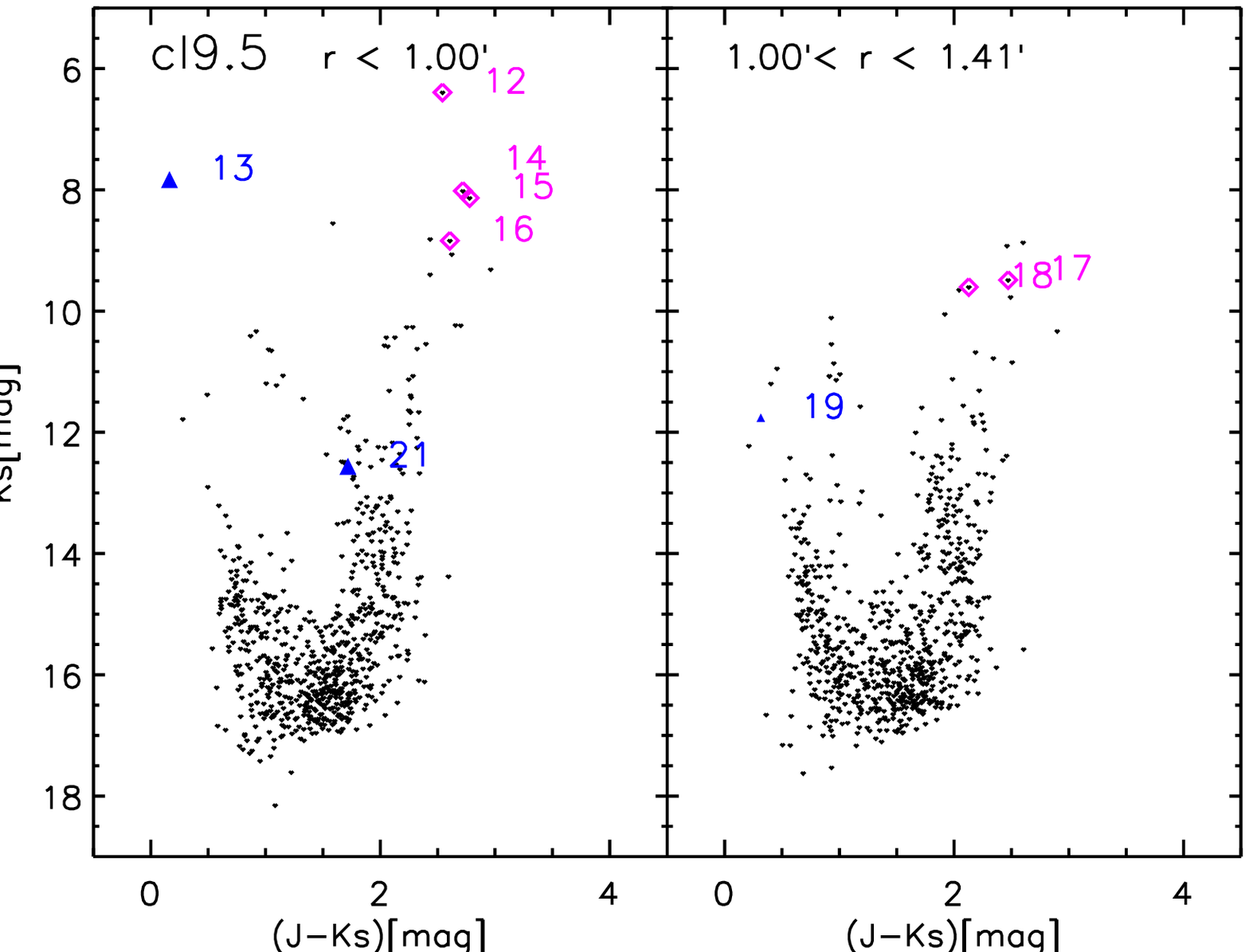}}
\resizebox{0.5\hsize}{!}{\includegraphics[angle=0]{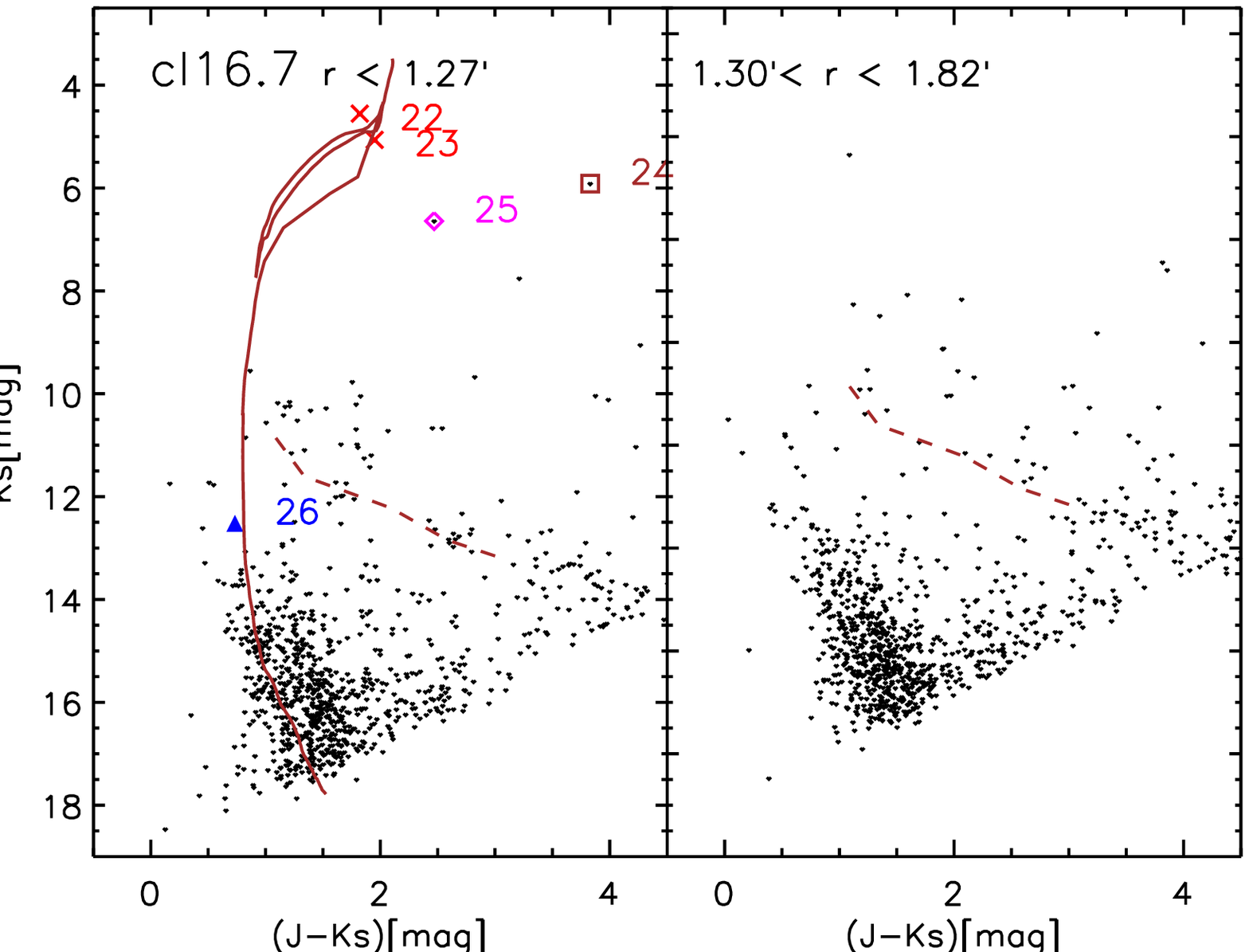}}
\resizebox{0.5\hsize}{!}{\includegraphics[angle=0]{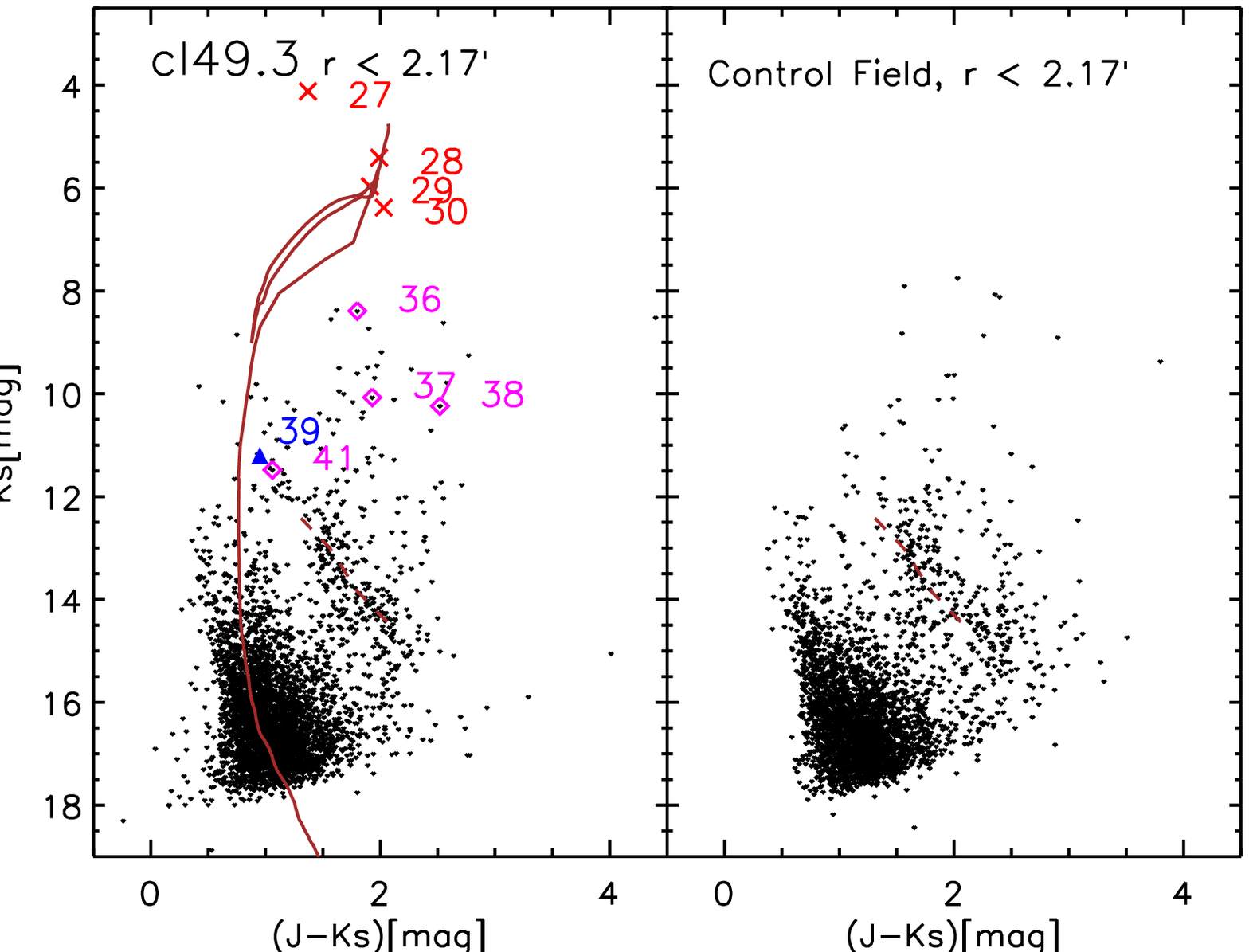}}
\resizebox{0.5\hsize}{!}{\includegraphics[angle=0]{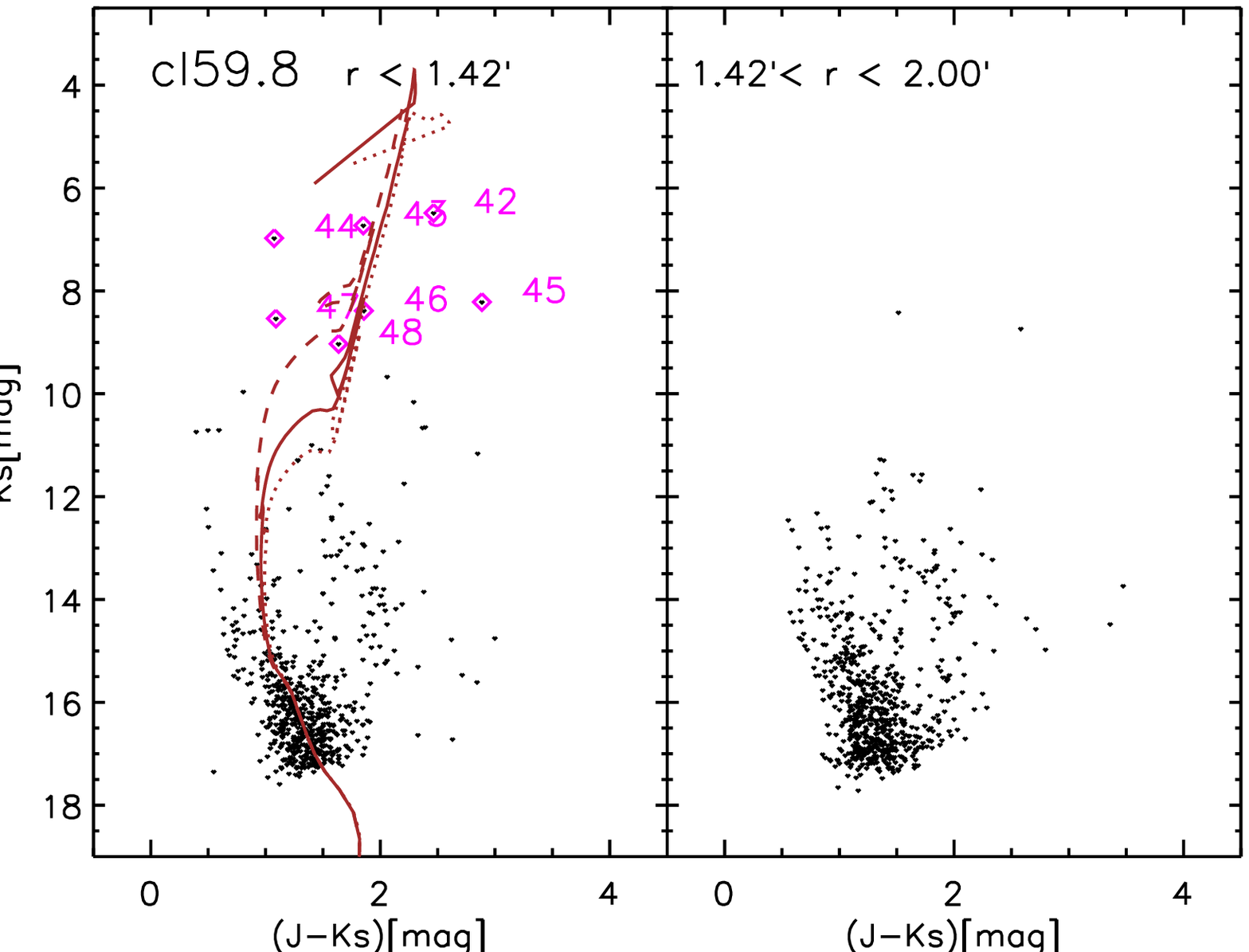}}
\end{centering}
\caption{\label{cmdcl1.5} 2MASS/UKIDSS $J-K_S$ versus $K_S$ CMDs: 
 cl1.5 ({\it top left}),  cl9.5 ({\it top right}),  cl16.3 ({\it middle left}), cl49.3 ({\it middle right}),
and  cl59.8 ({\it bottom left}). Photometry was obtained with a PSF-fitting technique (see text).
For each candidate  cluster,  a diagram of the  candidate cluster is shown in the left panel, and that of
a control field of equal area  in the right panel.  Spectroscopically  detected early and 
yellow stars are marked  with filled triangles, Miras with squares, RSGs with crosses, 
and red giants with diamonds;  identification numbers are taken from Table  \ref{table.phot}.
 In the CMDs of fields cl16.3  and cl49.3, which are rich in RSGs, 
a continuous curve displays an  isochrone of 20 Myr 
from \citet{lejeune01}, reddened with \Aks=0.50 mag and 0.48 mag, and shifted to DM= 12.93 mag and 14.22 mag, respectively;
a dashed line shows the sequence of red clump stars  (see Sect.\ \ref{secdist}).
The CMD of cl59.8 shows  three isochrones from \citet{marigo08}  with solar 
metallicity and  ages of 100 (dashed line), 200 (continuous line), and 300 Myr (dotted line), 
which were reddened (\Aks=0.55 mag) and shifted to DM=12.9 mag.}
\end{figure*}

The CMDs of the analyzed candidate clusters  are shown in Fig.\ \ref{cmdcl1.5}.
The CMDs of fields cl16.7 and cl49.3 are characterized by  gaps; in these two fields  we detected
several stars with luminosities typical of RSGs. The infrared CMDs nicely complement and strengthen 
the detection of six new RSGs, as illustrated in the following.

\subsection{RSGs in field cl16.7}

The 2MASS/UKIDSS ($J-$\Ks) versus \Ks\ diagram of field cl16.7 shows
two bright RSGs (stars 22 and 23) at ($J-$\Ks)$\approx 1.9$ mag and \Ks$\approx 4.8$ mag 
(Fig.\ \ref{cmdcl1.5}).
The two RSGs (stars 22  and 23) have similar extinction values (\Aks=0.46 and 0.52 mag), 
and spectral types,  and so  they are drawn from the same population.
The CMD shows a group of bright stars that extends over a main sequence to about  $9.5$ mag.
There is a gap of about 4.5 mag between these stars and the two RSGs.
The two other bright stars  with \Ks=5.91 and 6.64 mag are  field giant stars.

Star 26 (early or yellow-type)  has \Aks$\sim0.58$ mag, which is consistent with the extinction
of the two RSGs. On the CMD, the star is located at 
(J-\Ks)=$ 1.0$ mag and \Ks=$12.1$ mag, over a main sequence  (from \Ks=18 mag to 14 mag).
It is probably not related to the young RSGs; for DM=12.9 mag, we obtain  \Mk=$-1.4$ mag 
 \citep[typical of old late-B giants, see, e.g.,][]{clampitt97}.

Star 24 is 5\arcsec\ from the  IRAS 18207$-$1441 source,  within its positional error 
ellipse. Its spectrum has strong water absorption. A few rare RSGs have shown  water 
absorption in $H$ band, e.g., My Cep  (M7I) in  NGC 7419 \citep{rayner09}.  My Cep, the 
brightest of the five RSGs in NGC7419, is a pulsating star, and shows OH and water masers 
\citep{benson90}. Star 24 can be a late giant (M8III) or a faint M3I with an integrated 
Mbol of $-4.97$ mag. Since the two RSGs (K5I, K5.5I) are bluer and brighter 
(Mbol=$-6.2$ and $-5.7$ mag) than star 24, we concluded that this was most likely an 
unrelated AGB star. Accurate radial velocities are needed to confirm the proposed scenario.
 
%use lejeune only for the color locus
A colored theoretical isochrone from \citet{lejeune01}, with increased mass loss, 
solar metallicity, and an age of 20 Myr, is overplotted on the CMD of field cl16.7 
in Fig.\ \ref{cmdcl1.5}. The isochrone is shifted to the assumed cluster distance, 
and reddened. It encompasses the RSGs, as well as the early-type giant star 26.

For the two RSGs, we derived luminosities of $2.3 \times 10^4$ and $1.5 \times 10^4$ \Lsun\
(see Table \ref{luminosity}).
In Fig.  \ref{fig.luminosity}, luminosities are plotted versus effective temperatures,
and compared to theoretical stellar tracks by \citet{ekstrom12}; we used   tracks at  solar metallicity 
that include rotation; the RSGs had  initial masses from 10 \Msun\ to 13 \Msun.

\subsection{RSGs in field  cl49.3}

In field cl49.3, we detected four RSGs (27, 28, 29, and 30).
The four RSGs have similar  \Aks (\Aks=$0.47\pm0.11$ mag), and spectral types (from K4 to M1.5);
their \Ks\ range  from 4.1 mag to  6.4 mag. The $J-$\Ks\ versus \Ks\ diagram 
presents a gap of about 2 mag between the RSGs and other fainter stars (Fig.\ \ref{cmdcl1.5}).

The O9.5-B3 star 39 is  4.8 mag fainter than the group of  RSGs, and has \Aks=0.56 mag.
The spectro-photometric properties of star 39
indicate  a giant  (DM=14.16 mag) or a dwarf  (DM=13.02 mag). 
The similarity of  reddening suggests  that the giant 39 and  the four RSGs  are
at the same distance, and could belong to the same coeval population; 
an isochrone of 20 Myr \citep{lejeune01}  encompass  both the RSGs and the 
colors and magnitudes of star 39.

\begin{figure}[!]
\begin{centering}
\resizebox{0.7\hsize}{!}{\includegraphics[angle=0]{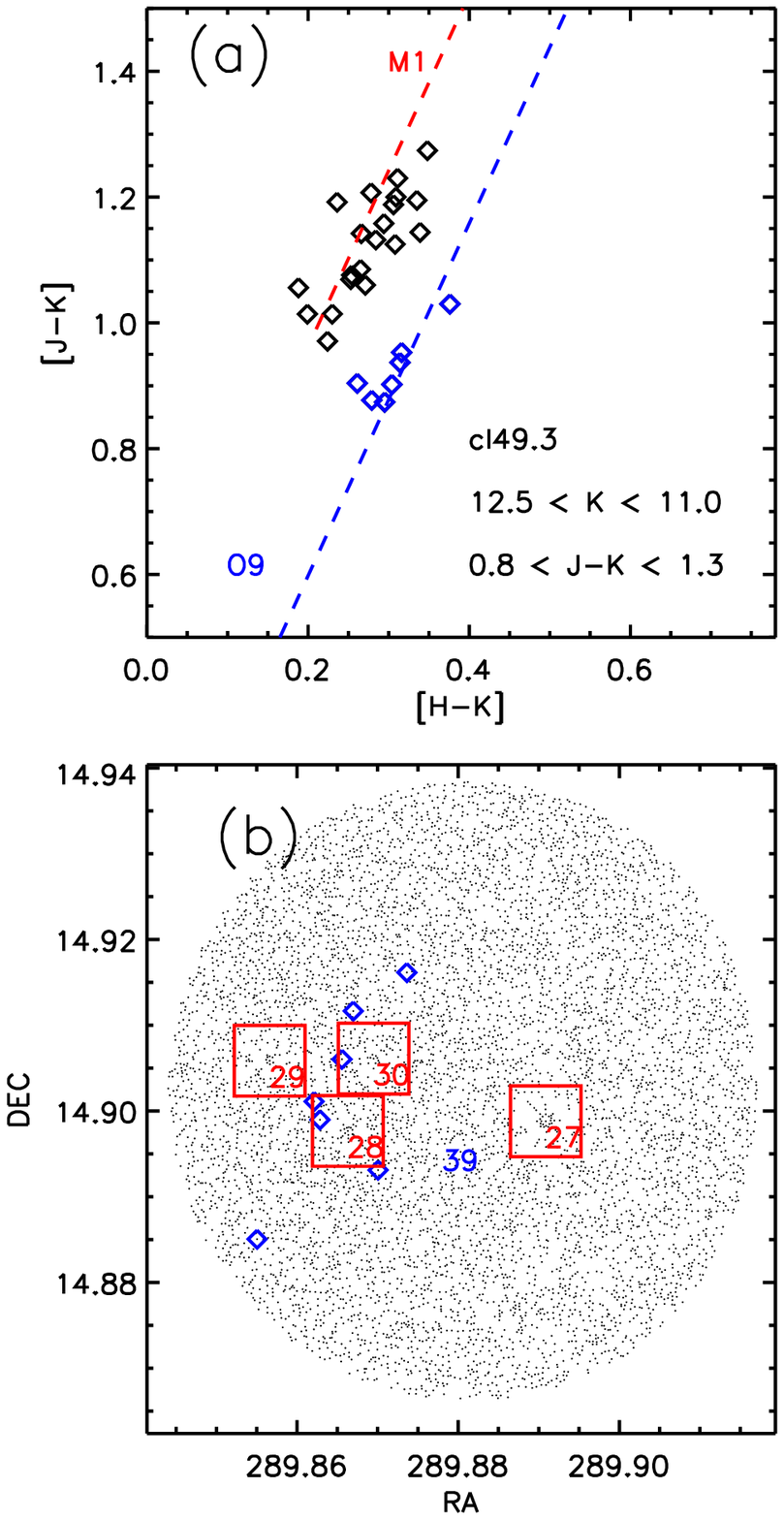}}
\end{centering}
\caption{\label{blue} 
 Panel (a) shows a ($J-K$) versus ($H-K$) diagram of stars with $12.5<K<11$ mag  and $ 0.8 <J-K <1.3$ mag 
in the cl49.3 field. The dashed lines show the locus of a 
M1 and an O9 star for increasing \Aks. 
Panel (b) shows} a map of the blue stars (diamonds), along with 
the detected RSGs (squares) and field stars (dots).
\end{figure}

The  bulk  of stars  at  $K=$11-12.5 mag and $J-K\approx1$ mag 
is made of red clump stars (see Sect. \ref{offthere}
and Fig.\ \ref{clump}). Among them we found  six  blue stars and star 39
 (see Fig.\ \ref{blue}).   These blue stars 
 appear radially concentrated towards the RSGs, and may be unrelated to the population of clump stars;
 late-types and early-types have  different intrinsic colors 
($J-K =0.7-0.8$ mag for late, and $\approx0.0$ mag for early-type stars), 
and the red clump of stars at $J-K\approx1$ mag is in the foreground of star 39.

\begin{figure}[!]
\begin{centering}
\resizebox{\hsize}{!}{\includegraphics[angle=0]{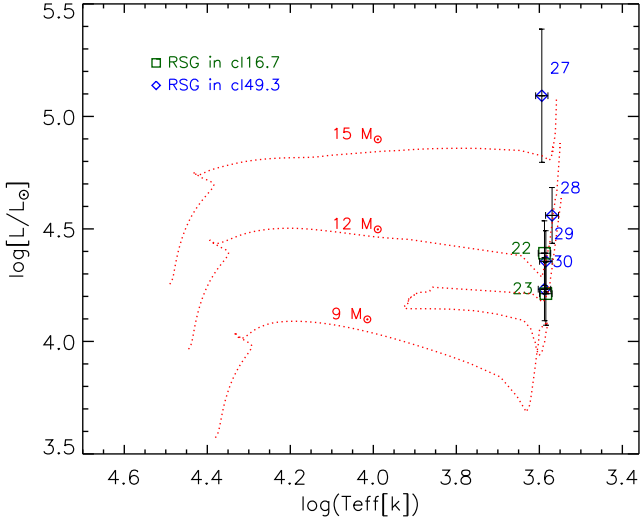}}
\end{centering}
\caption{\label{fig.luminosity}  
Luminosities versus effective temperatures of the newly detected RSGs.
RSGs in field cl16.7 are marked with squares, RSGs in field cl49.3 with diamonds.
Stellar tracks for stars of  9,  12, and 15 \Msun,
from the new rotating Geneva models with a solar metallicity \citep{ekstrom12}, 
are  shown with dotted  lines.}
\end{figure}

For the assumed distance (DM=$14.2\pm0.1$ mag from clump stars), the  luminosities of RSGs range 
from $1.6 \times 10^4$ \Lsun\ to $1.1 \times 10^5$ \Lsun.
From a comparison with the tracks by \citet{ekstrom12}, we estimate 
initial masses from 12 \Msun\ to 15 \Msun\ (see Fig.\  \ref{fig.luminosity}).

\subsection{ Other fields }

Field cl59.8 contains seven bright stars with \Ks\ from 6.5 mag to 9 mag.
For the estimated DM= $12.9\pm1.4$ mag, these stars 
have typical bolometric magnitudes of giant stars.  Nevertheless, they 
form a statistically significant 
overdensity of bright stars. The CMD of field cl59.8 is shown in Fig.\ 
\ref{cmdcl1.5}.
Stars 44 and 47 are bluer (\Aks=0.06 and 0.11 mag), and  are likely in the foreground.
Star 45 has \Aks=0.79 mag, and is  a possible background object.   
For the remaining four stars,  42,  43,
46, and 48, \Aks\ range from 0.36 mag to 0.56 mag, and
luminosities increase  with  spectral types;
they are consistent with massive giants drawn from a population older than 100 Myr \citep{marigo08}.
The infrared CMDs of simple stellar populations from 50 Myr to 200 Myr show
gaps and a large number of bright giants, e.g., the GLIMPSE13 cluster  \citep{messineo09}.

The CMDs of fields cl1.5 and cl9.5 do not present peculiar gaps or
overdensities of stars brighter than \Ks=7 mag (see Table \ref{counts}).
In conclusion,  fields cl1.5 and cl9.5  are populated by older 
stars \citep[$>$ 600 Myr;][]{ferraro95}.

Field cl1.5  appears  as a window of   low extinction  at the distance of the Galactic center
with an inferred DM=$14.54\pm0.16$ mag, and an average \Aks=$0.45\pm0.08$ mag;
this result is in agreement with the  extinction map by
\citet[ ][]{gonzalez12a}.
Spectroscopically detected giant stars  within $<2\arcmin$ from the center of cl1.5 
have \Ks\ from  7.52 mag  to 8.55 mag;  since   the tip of the
red giant branch (RGB) is  expected at $K \approx 8.2$ mag \citep{messineo05,glass85},
 they are most likely AGBs.
The Mira-like  star 2  is above the RGB tip (\Ks=6.86 mag). 
The Mira-like AGB 1 is $3\rlap{.}{^\prime}3$ from the center, and 
has \Ks=4.86 mag and  \Mbol$=-6.98$ mag for the assumed  DM, 
which is compatible with the maximum  luminosities
predicted for Mira stars \citep[see Fig.\ 20 in ][]{wood93}.

\subsection{The $Q1$ and $Q2$ parameters}
\citet{messineo12} analyzed Galactic stars of known types with the following parameters:

$Q1 = (J-H)-1.8 \times (H-$\Ks),\\

$Q2 = (J-$\Ks$)-2.69\times$(\Ks$-[8.0]).$ \\

These two parameters are interstellar extinction free, and measure the distance
of a datapoint from the Galactic reddening vector (passing trough the origin) in the $J-H$ versus $H-$\Ks\ ($Q1$), 
and $J-$\Ks\ versus \Ks$-[8.0]$ planes ($Q2$).
Typically, early-type stars have $Q1 < -0.20 \times (K_S-[8.0])+0.34$ mag \citep{messineo12}.
A fraction of 72\% of the spectroscopic targets are correctly
classified into late- and early-type stars by applying the above relation.
The $Q1$ and $Q2$ parameters only allow the selection of  stars with colors typical of  RSGs. 
Additional information on distances and luminosities  is needed to  classify RSGs.
\citet{messineo12} calculated that  72\% of known RSGs
fall into the region  $0.1 $ mag $< Q1 < +0.5$ mag and $-1.1$ mag $ < Q2 < +1.5$ mag;
five of the six new RSGs fall in this region, their average $Q1$=0.3 mag
with $\sigma$=0.12 mag; in contrast, detected Mira-like stars
have an average $Q1=-0.11$ mag with $\sigma$=0.14 mag.

\section{Summary and discussion}
\label{secsummary}

The detection of clusters rich in RSGs is strongly affected by biases.
It is difficult to classify and detect RSGs, because of our poor knowledge on
distance, and because of the high and patchy extinction in direction of the Galactic plane.
Nevertheless, their census is of importance to understanding the history of Galactic
chemical enrichment,  modality of star formation,  and Galactic structure.

We conducted infrared spectroscopic observations of five  overdensities
with bright stars, which were detected with GLIMPSE and 2MASS data along the Galactic plane.
We obtained 48 stellar spectra, and assigned  spectral types to  47 new stars.

Late-type stars were classified using photometric data and spectroscopic indexes
(water indexes and EW(CO)s).  High values of the water index showed that stars 1, 2, 
and 24 were Mira-like AGBs.
We detected six new RSGs (22, 23, 27, 28, 29, and 30)
via  their large EW(CO)s, low water indexes, and luminosities ( $> 10^4$ \Lsun).
All but one of the RSGs have EW(CO) $> 42$\AA.

Clustering of candidate RSGs and gaps in the CMDs are powerful tools for detecting RSGs. 
The detected RSGs are  in two fields, cl16.7 and cl49.3. 
Each group of RSGs presents narrow ranges of \Aks\ and  \Ks\ magnitudes.
Stars 22 and 23 fall in field cl16.7,  and stars 27, 28, 29, and 30 in 
field cl49.3. The CMDs of both fields show large gaps between the RSGs and the group
of blue giant/supergiants extending over a  main sequence; 
this gap is a typical feature of clusters rich in RSGs \citep[e.g.,][]{figer06}.

The two RSGs in field cl16.7 are located at a heliocentric distance of 
$\sim 3.9$ kpc, at a Galactic longitude of 16.7\degr\ on the the Scutum-Crux arm.
The two RSGs had stellar initial masses from 11 \Msun\ to 13 \Msun.

The cluster of four RSGs in field cl49.3 is at a  distance of $\sim 7.0$ kpc, 
at 49$\rlap{.}^{\circ}$3 of galactic longitude  on the Sagittarius-Carina arm. 
The RSGs had initial masses of 12 \Msun\ to 15 \Msun.
 
By using multiwavelength analysis, and a measure of  dust obscuration with distance,
it is possible to 
isolate far distant  and luminous objects, and groups of candidate RSGs. 
However,  spectroscopic confirmation is required.
Follow-up kinematic and variability studies will provide further
constraints on the properties of the
newly detected groups/clusters of  RSGs.
The $V$ band magnitudes of the newly discovered RSGs range from 13.7 mag to 17.2 mag, within the  
photometric and spectroscopic range covered by the upcoming mission Gaia. 
  
\begin{acknowledgements}
This work was partially funded by the ERC Advanced Investigator Grant GLOSTAR (247078).
The material in this work was partly supported by NASA under award NNG 05-GC37G, through 
the Long-Term Space Astrophysics program. 
This research
was partly performed in the Rochester Imaging Detector Laboratory
with support from a NYSTAR Faculty Development Program
grant.
 Qingfeng Zhu thanks for the support by the Fundamental Research Funds for the Central 
Universities from Educational Department of China, \#WK2030220001.
MM and VI thank ESO for supporting a working visit of Dr. Ivanov at  RIT.
MM and ZQ acknowledge an awarded support from ESA/ESTEC to host Dr. Zhu.
This publication makes use of data products from the Two Micron All Sky Survey, which is a joint project 
of the University of Massachusetts and the Infrared Processing and Analysis Center/California Institute of Technology, 
funded by the National Aeronautics and Space Administration and the National Science Foundation.
This work is based [in part] on observations made with the Spitzer Space Telescope, 
which is operated by the Jet Propulsion Laboratory, California Institute of Technology under a contract with NASA.
DENIS is the result of a joint effort involving human and financial
contributions  of several Institutes mostly  located in Europe. It has been
supported  financially mainly  by the French Institut National des Sciences de
l'Univers, CNRS, and French Education Ministry, 
the  European Southern Observatory, 
the State of Baden-Wuerttemberg, and  the European Commission under a  
network of the Human Capital and Mobility program.
This research made use of data products from the
Midcourse Space Experiment, the processing of which was funded by the Ballistic Missile Defence Organization with additional
support from the NASA office of Space Science. This research has made use of the SIMBAD data base, operated at CDS, Strasbourg,
France. This publication makes use of data products from
WISE, which is a joint project of the University of California, Los
Angeles, and the Jet Propulsion Laboratory/California Insti-
tute of Technology, funded by the National Aeronautics and
Space Administration. 
We are thankful to  the UKIRT team and SofI team for a great support during the observing runs,
to the  UKIDSS team in Cambridge for  providing detailed information on the pipeline products,
and to Dr. Stetson for providing the Daophot code, and helping with the installation.
MM thanks  Harm Habing and Ed Churchwell for discussions on GLIMPSE data,
and J. Borissova, R. Kurtev, Saurabh Sharma, and G. R. Ivanov
for discussions on stellar overdensities.
MM is grateful to  Jos de Bruine and  Timo Prusti for 
useful discussions and support during her ESA fellowship.
We thank the referee Ignacio Negueruela 
for his careful reading of our
manuscript.
\end{acknowledgements}

\end{document}